\documentclass[useAMS,usenatbib,usegraphicx]{mn2e}
\usepackage{color}

\newcommand\kms{{\rm\, km\, s^{-1}}}
\newcommand\pc{{\rm\, pc}}
\newcommand\cs{c_s}
\newcommand\F{\mathcal{F}}

\newcommand\Mperp{\mathcal{M}_\perp}

\newcommand\Sigp{\Sigma_{\rm peak}}
\newcommand\Mdot{\dot{M}}
\newcommand\Mtot{\dot{M}_{\rm tot}}
\newcommand\Mext{\dot{M}_{\rm ext}}
\newcommand\Mself{\dot{M}_{\rm self}}
\newcommand\Mshock{\dot{M}_{\rm shock}}

\newcommand\aMtot{\langle\dot{M}_{\rm tot}\rangle}
\newcommand\aMext{\langle\dot{M}_{\rm ext}\rangle}
\newcommand\aMself{\langle\dot{M}_{\rm self}\rangle}
\newcommand\aMshock{\langle\dot{M}_{\rm shock}\rangle}
\newcommand\Pext{\Phi_{\rm ext}}
\newcommand\Psp{\Phi_{\rm sp}}
\newcommand\Pgas{\Phi_{\rm gas}}

\newcommand\Omb{\Omega_b}
\newcommand\Omp{\Omega_p}
\newcommand\Oms{\Omega_p}

\newcommand\RCR{R_{\rm CR}}
\newcommand\RILR{R_{\rm ILR}}
\newcommand\RULR{R_{\rm 4/1}}
\newcommand\Rt{R_{\rm term}}

\newcommand\kpc{{\rm\, kpc}}
\newcommand\Mpc{{\rm\, Mpc}}
\newcommand\yr{{\rm\, yr}}
\newcommand\Gyr{{\rm\, Gyr}}

\newcommand\Vlos{V_{\rm LOS}}

\newcommand\Msun{{\rm\, M}_{\sun}}

\newcommand\freq{{\rm\, km\, s^{-1}\, kpc^{-1}}}

\newcommand\Surf{\Msun\;\rm pc^{-2}}
\newcommand\Aunit{\Msun \yr^{-1}}
\newcommand\simgt{\lower.5ex\hbox{$\; \buildrel > \over \sim \;$}}
\newcommand\simlt{\lower.5ex\hbox{$\; \buildrel < \over \sim \;$}}
\newcommand\HII{H{\scriptsize II}\,\,}

\title[Spiral Structure and Mass Drift]
{Gaseous Spiral Structure and Mass Drift in Spiral Galaxies}
\author[Kim and Kim]{Yonghwi Kim$^{1}$\thanks{e-mail : kimyh@astro.snu.ac.kr}
and Woong-Tae Kim$^{1,2}$\thanks{e-mail : wkim@astro.snu.ac.kr}\\
$^{1}$Center for the Exploration of the Origin of the Universe
(CEOU), Astronomy Program, Department of Physics \& Astronomy, \\
Seoul National University, Seoul 151-742, Republic of Korea \\
$^{2}$Center for Theoretical Physics (CTP), Seoul National
University, Seoul 151-742, Republic of Korea}
\begin{document}

\date{Accepted for publication in the MNRAS.}

\pagerange{\pageref{firstpage}--\pageref{lastpage}} \pubyear{2013}

\maketitle

\label{firstpage}

\begin{abstract}
We use hydrodynamic simulations to investigate nonlinear gas
responses to an imposed stellar spiral potential in disk galaxies.
The gaseous medium is assumed to be infinitesimally thin,
isothermal, and unmagnetized. We consider various spiral-arm models
with differing strength and pattern speed. We find that the extent
and shapes of gaseous arms as well as the related mass drift rate
depend rather sensitively on the arm pattern speed. In models where
the arm pattern is rotating slow, the gaseous arms extend across the
corotation resonance (CR) all the way to the outer boundary, with a
pitch angle slightly smaller than that of the stellar counterpart.
In models with a fast rotating pattern, on the other hand, spiral
shocks are much more tightly wound than the stellar arms, and cease
to exist in the regions near and outside the CR where $\Mperp/\sin
p_* \simgt 25$--40, with $\Mperp$ denoting the perpendicular Mach
number of a rotating gas relative to the arms with pitch angle
$p_*$. Inside the CR, the arms drive mass inflows at a rate of $\sim
0.05$--$3.0\Aunit$ to the central region, with larger values
corresponding to stronger and slower arms. The contribution of the
shock dissipation, external torque, and self-gravitational torque to
the mass inflow is roughly 50\%, 40\%, and 10\%, respectively. We
demonstrate that the distributions of line-of-sight velocities and
spiral-arm densities can be a useful diagnostic tool to distinguish
if the spiral pattern is rotating fast or slow.
\end{abstract}

\begin{keywords}
galaxies: ISM -- galaxies: kinematics and dynamics --
galaxies: structure -- galaxies: nuclei -- galaxies: spiral  --
hydrodynamics -- ISM: general -- shock waves
\end{keywords}

\section{Introduction}

Disk galaxies possess prominent non-axisymmetric features such as
spiral arms and bars that have profound influences on galactic
evolution in various ways (e.g., \citealt{but96,kor04,but13,sel13}
and references therein). Stellar spiral arms not only trigger or
organize star formation in the outer parts of disk galaxies but also
drive secular changes in the orbits of stars and gas clouds,
redistributing the mass in the disks (e.g.,
\citealt{lin64,lin66,too64,elm95,ber96,foy10}). Bar potentials also
affect the mass redistributions in the inner parts and are
responsible for the formation of various gaseous substructures such
as dust lanes and nuclear rings (e.g.,
\citealt{san76,ath92,hel94,pin95,but96,kim12a}). Understanding the
gravitational interaction of the stellar potentials with a gaseous
medium is therefore the first step to understand star formation,
secular evolution, and morphological changes occurring in disk
galaxies.

Among various secular processes, an angular momentum exchange
between gas and a stellar pattern is particularly interesting since
it leads to overall gas inflows or outflows in the radial direction.
In barred galaxies, it has been well established that a
non-axisymmetric torque exerted by a bar potential produces a pair
of dust lanes in the gaseous medium, across which the gas loses
angular momentum and falls radially inward to form a nuclear ring
(e.g.,
\citealt{ath92,pin95,eng97,pat00,mac04,ann05,tha09,kim12a,kim12}),
potentially powering star burst activities in nuclear rings as well
as fueling active galactic nuclei (e.g.
\citealt{shl90,reg99,kna00,lau04,jog05,hun08,van10}). In particular,
\citet{kim12b} demonstrated that the location of nuclear rings is
determined not by the resonances but by the centrifugal barrier that
the inflowing gas cannot overcome.  This suggests that nuclear rings
are smaller in size in more strongly-barred galaxies, entirely
consistent with the observational results of \citet{com10}.

Compared to the cases with a bar potential, the effects of a spiral
potential on the gaseous structures and radial mass inflow are
relatively poorly understood. While interactions of stellar density
waves with stars and the associated stellar heating and radial
migration have been a subject of intense study (e.g.,
\citealt{lyn72,don94,zha96,ath02,sel02,ros08,oh08,sel11,bru11}; see
also \citealt{kor04} and \citealt{sel13} and references therein),
only a few studies have explored angular momentum loss of gas due to
the density waves (e.g., \citealt{kal72,rob72,lub86,hop11}). For
example, \citet{rob72} considered non-self-gravitating galactic
spiral shocks and showed that the damping timescale of stellar
density waves due to the angular momentum exchange is of the order
of $\sim1\Gyr$. \citet{lub86} included the back reaction of the
density waves on the gas density distribution, finding that the gas
accretion rate due to the stellar pattern amounts to $\Mdot\sim
-(0.2$--$0.4)\Aunit$ for parameters representing the solar
neighborhood in the Milky Way. These values of $\Mdot$ are overall
consistent with the mass inflows inferred from chemical modeling in
the Milky Way \citep{lac85} and also those in external galaxies
based on gravitational torque analyses \citep{haa09,gar09}.
\citet{hop11} used the epicycle approximation to derive an analytic
expression for $\Mdot$ due to a non-axisymmetric potential.

While the derivations of $\Mdot$ by \citet{lub86} and \citet{hop11}
are insightful, they utilized a few notable approximations. First,
\citet{lub86} considered local, tightly-wound waves in both the
stellar and gaseous media and ignored the self-gravitational torque
on the gas in evaluating $\Mdot$. They also included shear viscosity
to represent cloud collisions, which tends to smear out shock
profiles and thus makes it difficult to isolate the sole effect of
the shock (e.g., \citealt{kim07,kim08}). On the other hand,
\citet{hop11} used the orbit crossing of test particles as a
criterion for the shock formation, without considering the effects
of gas pressure as well as the speed of incident flows relative to
the pattern. Although \citet{hop11} showed that their $\Mdot$ is in
good agreement with their numerical results for galaxies with a
dominant bar-like potential, it is uncertain whether the same holds
true for spiral galaxies in which the effects of thermal and ram
pressures are more important than in barred galaxies.

Since dynamics involved with spiral arms is intrinsically nonlinear,
it is desirable to run numerical hydrodynamic simulations in order
to measure the mass drift rates driven by spiral arms properly.
There have been numerous studies for gas responses to an imposed
spiral potential, focusing on morphological changes of gaseous arms
depending on the pattern speed (e.g.,
\citealt{pat94,pat97,gom02,sly03,yan08,gom13}), formation of arm
substructures such as branches, spurs, and feathers (e.g.,
\citealt{cha03,wad04,she06,dob06}; see also local models of
\citealt{kim02,kim06}), or star formation occurring in spiral arms
(e.g., \citealt{she08,wad08,wad11,dob11}). While the arm-driven mass
inflows might have affected the simulation outcomes in the work
mentioned above, its rate has yet to be evaluated to assess its
dynamical consequences on secular evolution quantitatively.

In this paper, we run global hydrodynamic simulations for gas
evolution in galaxies with spiral potentials. We consider an
infinitesimally-thin, uniform gaseous disk.  We take an isothermal
equation of state for the gas, and ignore the effect of radiative
cooling and heating, star formation, and magnetic fields. Two
important parameters characterizing a spiral pattern are its angular
frequency $\Oms$ and strength $\F$, which are difficult to constrain
observationally. Thus we in this work vary $\Oms$ and $\F$ as free
parameters to model spiral arms in various galactic situations, and
study how $\Mdot$ depends on them. We will also compare our
numerical results with the analytic expression presented in
\citet{hop11}.

In addition to evaluating $\Mdot$, our models are also useful to
address important issues related to the spatial extent, structures,
and pitch angles of gaseous arms in comparison with their stellar
counterparts. While the theory for spiral density waves suggests
that the \emph{stellar} pattern extends up to the corotation
resonance (CR) or to the outer Lindblad resonance (OLR) if it is in
the linear-regime (e.g., \citealt{too81,lin79,ber89a,ber89b,zha96})
and to the 4/1 resonance if it is strong enough to be nonlinear
\citep{con86,con88,pat91}, it is uncertain whether the termination
of \emph{gaseous} arms corresponds to the resonance radii. Moreover,
\citet{git04} showed using both semi-analytic and numerical
approaches that gaseous arms are in general more tightly wound than
the stellar arms. Although large uncertainties surround
observational determinations of arm pitch angles, recent studies
show that they are, statistically, slightly larger in the $I$- or
$H$-band than in the $B$-band \citep{sei06,dav12,mart12}, suggesting
that the gaseous arms are likely more tightly wound than the stellar
counterpart. We will show that the extent and pitch angles of
gaseous arms are dependent somewhat sensitively upon the arm pattern
speed. We will also show that the distributions of line-of-sight
velocities in the projected galactic disk and density profiles of
gaseous arms can be used to tell whether the observed arms are
inside their CR or not.

The remainder of this paper is organized as follows. In Section
\ref{sec:model}, we describe the galaxy model and our choices of the
model parameters, as well as the numerical method we use. In Section
\ref{sec:arm}, we present the simulation results on morphologies of
spiral shocks. In Section \ref{sec:mdot}, we measure the arm-induced
mass drift rates as functions of the pattern speed and strength of
the stellar arms. In Section \ref{sec:los}, we present the
distributions of line-of-sight velocities in the plane of sky and
discuss how they can be used to obtain information on the arm
pattern speed. In Section \ref{sec:sum_dis}, we conclude with a
summary and discussion of our results and their astronomical
implications.

\section{Model and Method}\label{sec:model}

We consider an infinitesimally-thin, self-gravitating gaseous disk,
and study its nonlinear responses to an imposed non-axisymmetric
potential representing stellar spiral arms. The disk is assumed to
be unmagnetized and isothermal, for simplicity. The basic equations
of hydrodynamics expanded in the $z=0$ plane corotating with the
spiral potential are
\begin{equation}\label{eq:con}
\frac{\partial \Sigma}{\partial t}+\nabla\cdot({\Sigma \mathbf{u}})= 0,
\end{equation}
\begin{eqnarray}
\frac{\partial \mathbf{u}}{\partial t}+(\mathbf{u}\cdot\nabla)
\mathbf{u}=-\frac{\cs^2}{\Sigma}\nabla \Sigma-\nabla (\Pext + \Pgas)\label{eq:mom}\\
+\,\Oms^2\mathbf{R}-2\mathbf{\Omega}_s\times\mathbf{u},\qquad\quad\,\nonumber
\end{eqnarray}
\begin{equation}\label{eq:grav}
\nabla^2\Pgas = 4\pi G f(z) \Sigma,
\end{equation}
where $\Sigma$ is the gas surface density, $\mathbf{u}$ is the
velocity in the rotating frame, $\cs$ is the isothermal speed of
sound, $\mathbf{\Omega}_p=\Oms\hat{z}$ is the pattern speed of the
spiral arms, and $\Pext$ and $\Pgas$ denote the external
gravitational potential and self-gravitational potential,
respectively. In equation (\ref{eq:grav}), the function $f(z)$ is
introduced to account for the dilution of self-gravity at the disk
mid-plane due to finite disk thickness: we take a Gaussian profile
with thickness of $H=0.1R$ \citep{she08}. The velocity $\mathbf{v}$
in the inertial frame is obtained from $\mathbf{v} = \mathbf{u} +
R\Omp\mathbf{\hat \phi}$.

\subsection{Galaxy Model}

\begin{figure}
\centering
\includegraphics[angle=0,width=0.48\textwidth]{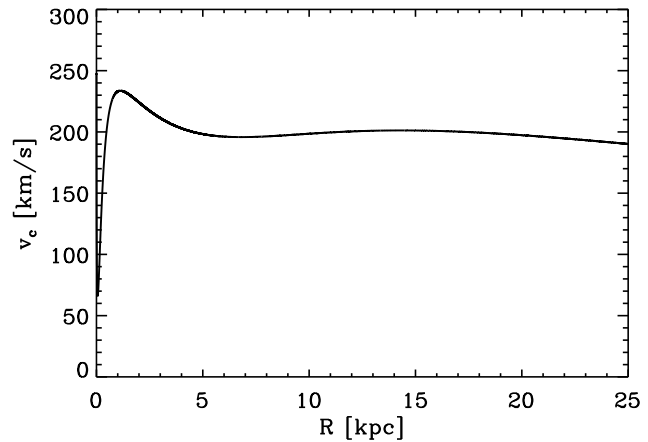}
\caption{Rotational velocity $v_c$ of our galaxy model as a function
of the galactocentric radius $R$. Over much of the disk, $v_c\simeq
200\kms$, corresponding to a normal disk galaxy.}\label{fig-rot}
\end{figure}

The external gravitational potential consists of an axisymmetric
component and a non-axisymmetric spiral component $\Psp$. The static
axisymmetric part responsible for galaxy rotation is comprised of a
stellar disk, a spherical bulge/halo, and a central black hole with
mass $M_{\rm BH}=4\times10^7\Msun$, identical to that in
\citet{kim12b}. Figure \ref{fig-rot} plots the resulting rotation
curve, with a flat part with $v_c\simeq 200 \kms$ over most of the
disk plane and a rapidly rising part as $v_c\propto (M_{\rm
BH}/R)^{1/2}$ toward the center due to the presence of the black
hole.

For the non-axisymmetric spiral potential, we take a
trailing logarithmic-arm model of \citet{she06}:
\begin{eqnarray}
\Psp(R,\phi;t)=\Phi_{0}\cos\left(m\left[\phi+ \frac{\ln R}{\tan
p_*}-\Oms t+\phi_0 \right]\right),\label{eq:sp1}
\end{eqnarray}
for $R\geq 2\kpc$ (see also \citealt{rob69}). Here, $m$, $p_*$,
$\Oms$, and $\phi_0$ denote the number, the pitch angle, the pattern
speed, and the initial phase of the arms, respectively. Note that
$\Psp$ is tapered from $R=2\kpc$ to $1\kpc$ by a Gaussian function
to have $\Psp=0$ at $R\leq 1\kpc$. The amplitude of the spiral
potential $\Phi_{0}$ in equation (\ref{eq:sp1}) is controlled by the
dimensionless parameter
\begin{equation}\label{eq:sp2}
\F\equiv \frac{m|\Phi_{0}|}{v_c^2 \tan p_*},
\end{equation}
which measures the gravitational force due to the spiral arms in the
direction perpendicular to the arms relative to the radial force
from the background axisymmetric potential (e.g.,
\citealt{rob69,kim02,kim06,oh08,she08}). We fix $m=2$,
$p_*=20\degr$, and $\phi_0=147\degr$. We vary $\F$ from 5 to 20\% to
study situations with differing arm strength.

It is quite challenging to measure the pattern speeds of spiral
arms. Self-consistent modeling of the Milky Way shows that stability
and observed tangent points of the spiral arms are best described by
$\Oms=20\freq$ (e.g., \citealt{ama97,mar04}). On the other hand,
analyses based on the \citet{tre84}'s method indicate that the arms
in external galaxies have pattern speeds in a wide range of
$\Oms\sim 10$--$45\freq$ (e.g., \citealt{zim04,fat09,mart11}).
$N$-body simulations for the formation of non-axisymmetric features
in disk galaxies also show that the angular frequency of spiral arms
is diverse (e.g., \citealt{sel88,rau99,bou02,roc13}). Therefore, we
in this paper consider three different cases: $\Oms=30\freq$ (fast
arm models), $\Oms =20\freq $ (intermediate-speed arm models), and
$\Oms=10\freq$ (slow arm models). In what follows, we refer to these
as the F, I, and S models, respectively. The positions of the CR,
the inner Lindblad resonance (ILR), and the 4/1 resonance are
$\RCR=6.5$, 9.9, and $19.8\kpc$, $\RILR=2.5$, 3.6, and $6.0\kpc$,
and $\RULR=4.4$, 6.4, and $12.9\kpc$ for the F, I, and S models,
respectively.

We run 18 models that differ in $\F$, $\Oms$, and the presence or
absence of gaseous self-gravity. Table \ref{tbl:model} lists the
model parameters. The prefixes ``F'', ``I'', and ``S'' stand for the
models with fast, intermediate-speed, and slow arms, while the
postfixes ``G'' and ``N'' indicate self-gravitating and
non-self-gravitating models, respectively. In all models, the gas
sound speed is taken to $\cs=10\kms$ that effectively includes a
contribution of turbulent motions (e.g., \citealt{mck07}). For
models in which self-gravity is considered, $\Pgas$ is calculated by
using the \citet{kal71} scheme described in \citet{she08}. All the
models start from a gaseous disk with uniform surface density
$\Sigma_0=10\Msun\pc^{-2}$. We take Models F10G, I10G, and S10G with
$\F=10\%$ as our fiducial models.

\begin{table}
\centering
\caption{Model Parameters\label{tbl:model}}
\label{symbols}
\begin{tabular}{cccc}
\hline
Model & $\F$ & $\Omega_s$ & Self-gravity \\
{} & (\%) & ($\freq$) & {} \\
\hline
F05G & 5  & 30 & included \\
F10G & 10 & 30 & included \\
F20G & 20 & 30 & included \\
I05G & 5  & 20 & included \\
I10G & 10 & 20 & included \\
I20G & 20 & 20 & included \\
S05G & 5  & 10 & included \\
S10G & 10 & 10 & included \\
S20G & 20 & 10 & included \\
\hline
F05N & 5  & 30 & omitted \\
F10N & 10 & 30 & omitted \\
F20N & 20 & 30 & omitted \\
I05N & 5  & 20 & omitted \\
I10N & 10 & 20 & omitted \\
I20N & 20 & 20 & omitted \\
S05N & 5  & 10 & omitted \\
S10N & 10 & 10 & omitted \\
S20N & 20 & 10 & omitted \\
\hline
\end{tabular}
\end{table}

\subsection{Numerical Methods}\label{sec:methods}

As in \citet{kim12b}, we integrate equations
(\ref{eq:con})--(\ref{eq:grav}) using the CMHOG code in cylindrical
geometry. CMHOG is a grid-based code for ideal hydrodynamics based
on the piecewise parabolic method in its Lagrangian remap
formulation \citep{col84}, which is third-order accurate in space
\citep{pin95}. All the simulations are performed in a frame
corotating with the arms. In the simulation domain, therefore, the
spiral potential remains stationary. In order to avoid strong
transients in the gas flows caused by a sudden introduction of the
spiral potential, we increase its amplitude slowly over the
timescale of $0.1\Gyr$.  We run the simulations until $t=1\Gyr$,
beyond which the numerical results are compromised by waves
propagating from the outer radial boundary.

By assuming a reflection symmetry with respect to the galaxy center,
the simulations are performed on a half-plane with
$-\pi/2\le\phi\le\pi/2$. We set up a logarithmically-spaced
cylindrical grid over $R = 0.5 \kpc$ to $40\kpc$, with 1102 radial
and 790 azimuthal grid points. The corresponding grid spacing is
$\Delta R = 2$, 40, and $159\pc$ at the inner radial boundary, at
$R=10\kpc$, and at the outer radial boundary, respectively. We apply
the continuous and outflow boundary conditions at the outer and
inner radial boundaries, respectively, while adopting the periodic
boundary conditions at the azimuthal boundaries. The gas moving in
through the inner radial boundary is considered lost out of the
simulation domain.

\section{Spiral Structures}\label{sec:arm}

In this section, we focus on spiral structures induced by the
imposed spiral potential. Radial mass flows associated with the
spiral shocks will be presented in Section \ref{sec:mdot}.

\subsection{Overall Morphology}\label{sec:overall}

\begin{figure*}
\centering
\includegraphics[angle=0,width=0.96\textwidth]{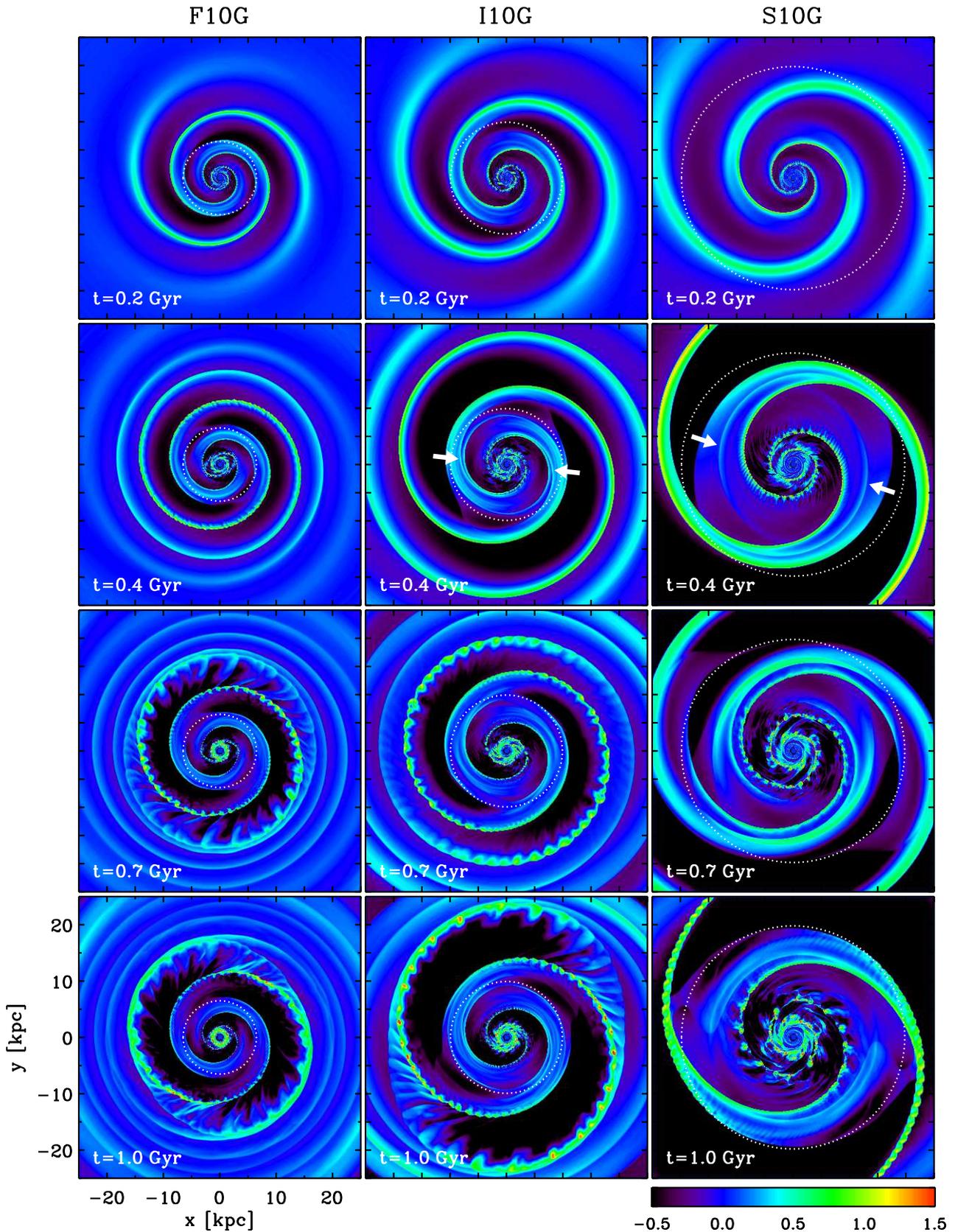}
\caption{Snapshots of gas surface density in logarithmic scale for
Models F10G, I10G, and S10G from left to right at $t=0.2$, 0.4, 0.7,
and $1.0\Gyr$ from top to bottom. In each panel, the CR of the arms
is indicated as a dotted circle. The white arrows in the $t=0.4\Gyr$
panels of Models I10G and S10G indicate weak structures emanating
from the 4/1 resonance. Colorbar labels
$\log(\Sigma/\Sigma_0)$.}\label{fig-sigt}
\end{figure*}

We begin by describing the temporal evolution of our fiducial models
with $\F=10\%$: the evolution of other models with different arm
strength is qualitatively similar. Figure \ref{fig-sigt} plots
snapshots of the gaseous surface density in logarithmic scale at
$t=0.2$, 0.4, 0.7, and $1.0\Gyr$ for the self-gravitating models.
The left, middle, and right columns are for the F, I, and S models
with $\Oms=30$, 20, and $10\freq$, respectively. The spiral arms
remain stationary in the simulation domain. The dotted circle in
each panel marks the CR of the spiral arms, outside of which the gas
is rotating in the clockwise direction relative to the spirals.

It is apparent that the spiral potential strongly perturbs the gas
orbits, forming large-scale spiral shocks, although the regions
affected by the potential depend on $\Oms$. In the F and I models,
spiral shocks are strong only inside the termination radius of
$\Rt\approx 17$ and $25\kpc$, respectively, outside which
small-amplitude perturbations propagating outward do not develop
into shocks.  On the other hand, the whole disk is strongly affected
by the spiral potential to induce shocks in the S models. As we will
show more quantitatively in Section \ref{sec:str} below, this is
because the gas does not have sufficient time to respond to a spiral
potential when it rotates too rapidly.  In the F and I models, the
outer ends of gaseous arms, which cannot be extended beyond $\Rt$,
curl back radially in and are loosely connected to the other arms at
late time, producing a ring-like structure just inside $\Rt$. These
ring-like structures are more vividly evident in models with large
$p_*$, as in models presented in \citet{pat94}. We defer a more
detailed discussion on this issue to Section \ref{sec:dis}.

\begin{figure*}
\centering
\includegraphics[angle=0,width=1.0\textwidth]{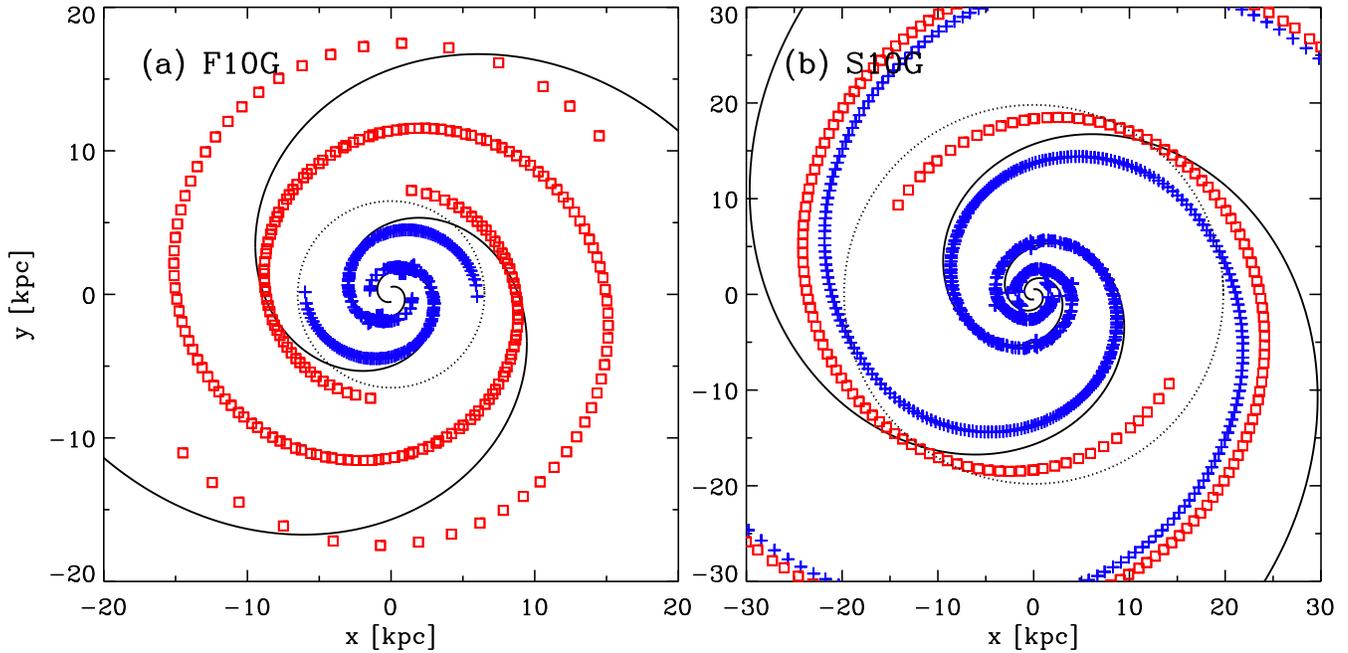}
\caption{Locations (symbols) of spiral shocks in Models F10G and
S10G at $t=0.4\Gyr$. Pluses and squares indicate the shocks produced
primarily by the gas inside and outside the CR, respectively. The
solid lines draw the loci of the spiral potential minima, while the
dotted circle marks the CR of the spiral pattern.} \label{fig-alpha}
\end{figure*}

Figure \ref{fig-sigt} shows that the density snapshots at
$t=0.4\Gyr$ of Models I10G and S10G contain weak gaseous structures,
indicated by the arrows, developing from the arms near the 4/1
resonance.\footnote{The 4/1 resonance corresponds to the first
ultraharmonic resonance for 2-armed spirals (e.g.,
\citealt{cha03}).} They are continually generated from the 4/1
resonance due to the nonlinear effects (e.g.,
\citealt{con86,con88,art92}), and propagate through the disk. These
weak structures share a lot of similarities in appearance and in
geometrical locations with ``branches'' and ``spurs'' identified by
\citet{cha03} (see also \citealt{pat94,pat97,yan08}).  These
terminologies of resonance features are visually motivated to
indicate structures bifurcating from the main arms. Branches refer
to trailing structures winding in the same sense as the main arms
such as in Model I10G, whereas those leading the arms such as in
Model S10G are termed spurs \citep{cha03}.  We find that models with
larger $\F$ and/or smaller $\Oms$ tend to produce spurs more easily,
while models with smaller $\F$ and/or larger $\Oms$ are more likely
to possess branches, consistent with the results of \citet{cha03}.
We note that bifurcations of gaseous arms in our models are much
weaker than those reported in \citet{pat94,pat97}, owing to a small
pitch angle. We have run additional models (not listed in Table
\ref{tbl:model}) with pitch angles of $p_*=33\degr$ and $44\degr$,
and confirmed that arms with a large pitch angle indeed develop
strong bifurcations. These results suggest that the growth of
resonance features is highly sensitive to the arm parameters.

Figure \ref{fig-sigt} also shows that some parts of spiral shocks
wiggle and form small clumps along them at late time. This
clump-forming wiggle instability is more virulent when shocks are
stronger. Yet, its physical nature is uncertain. Based on the
Richardson criterion, \citet{wad04} argued that it is the
Kelvin-Helmholtz instability of a shear layer behind the shock,
although the expanding radial velocity after the shock has a local
stabilizing effect \citep{dwa96}. \citet{dob06} interpreted the
clump formation as orbit crowding of gaseous particles that change
their angular momenta in the shock. \citet{kim12a} suggested that
vorticity at the curved shocks increases secularly due to Crocco's
theorem, producing clumps in the nonlinear stage. On the other hand,
it cannot be ruled out the possibility that the wiggle instability
can be of a numerical origin, caused by the inability of a numerical
method to properly resolve a shock inclined to numerical grids
\citep{han12}. As Figure \ref{fig-sigt} shows, the wiggle
instability develops only in the regions well away from the CR,
since its growth requires strong shocks \citep{wad04,kim06}. The
wiggle instability is weaker in models with a lower pattern speed,
which is most likely due to the fact that large gas velocities
relative to the spiral potential in the simulation domain have
considerable numerical viscosity (e.g., \citealt{kim08}), tending to
suppress the wiggle instability.\footnote{We find that spiral shocks
in models with parameters identical to those of Models F10G and S10G
calculated in the inertial frame rather than in the rotating frame
are stable to the wiggle instability, indirectly demonstrating
stabilization by numerical viscosity.} We note that regardless of
its nature, the wiggle instability occurring on small scale does not
significantly affect the radial gas drift rate averaged both
azimuthally and temporally.

Gas self-gravity does not make significant differences in the
overall arm morphologies, since our initial disk has the Toomre
stability parameter
\begin{equation}\label{eq:toomre}
Q_T=\frac{\kappa c_s}{\pi G\Sigma} \approx
2.1\left(\frac{R}{10\kpc}\right)^{-1}
\left(\frac{\Sigma}{10\Msun\pc^{-2}}\right)^{-1},
\end{equation}
which is larger than unity at $R\simlt 20\kpc$. Here $\kappa$
denotes the epicycle frequency. Although $Q_T <1$ in the outer
regions, Figure \ref{fig-sigt} shows that gravitational instability
does not manifest itself presumably due to the effect of finite disk
thickness in solving the Poisson equation (\ref{eq:grav}).

\subsection{Spiral Shocks}\label{sec:shock}

\subsubsection{Structure}\label{sec:str}

A global stellar spiral pattern that persists throughout the disk
perturbs gas orbits that would otherwise remain circular, and
produces spiral shocks in the gas flows. To measure the strength of
spiral shocks, we define the dimensionless compression factor
\begin{equation}\label{eq:alpha}
\alpha\equiv -(\nabla\cdot\mathbf{v})\Delta R/c_s\,,
\end{equation}
(e.g., \citealt{mac04,tha09,kim12a}). For steady isothermal shocks
in planar geometry, $\alpha = \Mperp - 1/\Mperp$, where $\Mperp$ is
the Mach number of the incident flows perpendicular to the shock
fronts. Thus, any positive value of $\alpha$ should correspond to
shocks if the flows are one-dimensional and in steady state.
However, the gas flows in the disk are two-dimensional and are not
completely steady. We empirically found that shock fronts (i.e.,
discontinuities in density and velocities) in our models are well
identified by the condition $\alpha\simgt 0.5$. In what follows, we
thus impose $\alpha\geq0.5$ as a condition for the presence of
spiral shocks.

\begin{figure}
\centering
\includegraphics[angle=0,width=0.48\textwidth]{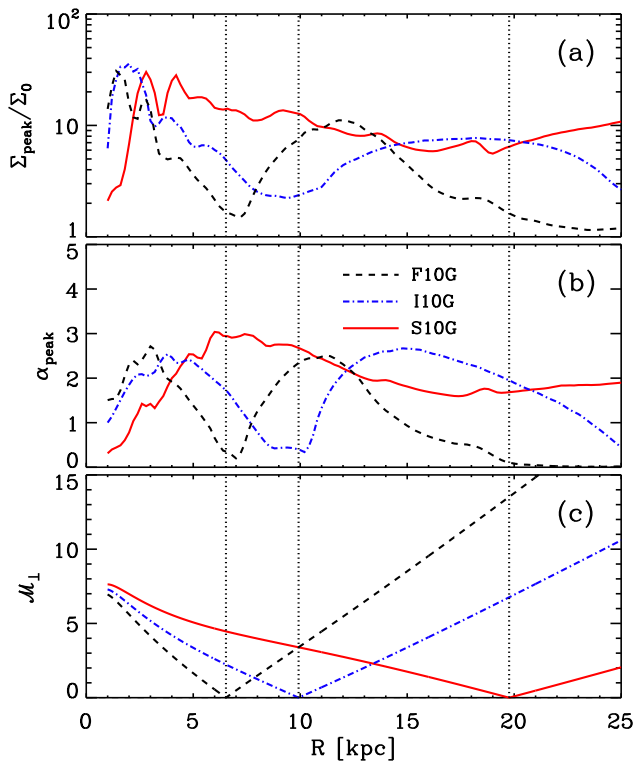}
\caption{Radial dependence of (a) the peak density $\Sigp/\Sigma_0$,
(b) the maximum compression factor $\alpha_{\rm peak}$, and (c) the
perpendicular Mach number $\Mperp$ for Models F10G (dashed), I10G
(dot-dashed), and S10G (solid) at $t=0.4\Gyr$. The vertical dotted
lines at $R=6.5\kpc$, $9.9\kpc$, and $19.8\kpc$ mark the CR of the
arms in the F, I, and S models, respectively.} \label{fig-mach}
\end{figure}
\begin{figure}
\centering
\includegraphics[angle=0,width=0.48\textwidth]{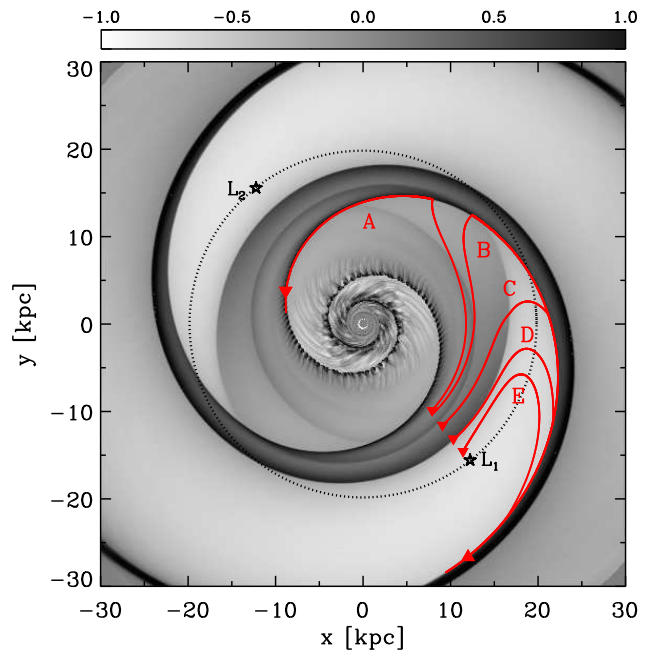}
\caption{A few instantaneous streamlines of the gas, starting from
near the Lagrangian point $L_1$, in Model S10G at $t=0.4\Gyr$ in the
frame corotating with the spiral potential. $L_2$ is another
Lagrangian point at the opposite side. The dotted circle denotes the
position of the CR. The streamline marked by ``A'' hits the shock at
an oblique angle less than $90\degr$ and bends radially inward
downstream, while the other streamlines meet the shocks at angles
larger than $90\degr$ and thus move radially outward after the
shocks. Grayscale bar labels $\log(\Sigma/\Sigma_0)$.
\label{fig-stm}}
\end{figure}

To delineate the positions and shapes of the spiral shocks, Figure
\ref{fig-alpha} plots as plus and square symbols the loci of the
maximum $\alpha$ $(\geq0.5)$ at each $R$ for Models F10G and S10G at
$t=0.4\Gyr$. In each panel, the dotted circle indicates the CR of
the arms, while the black solid curves mark the loci of the spiral
potential minima. It is apparent that spiral shocks in Model S10G
are located relatively close (within $\sim 20\degr$ in $\phi$) to
the potential minima, while they do not follow the spiral potential
closely in Model F10G. Figure \ref{fig-mach} plots the radial
variations over $1\kpc\leq R \leq 25\kpc$ of the peak density
$\Sigp$, the peak compression factor $\alpha_{\rm peak}$, and the
perpendicular Mach number $\Mperp = R|\Omega-\Oms|\sin p_*/\cs$ of
the incident flows relative to the arms for our standard models at
$t=0.4\Gyr$. The vertical dotted lines mark the CR of the arms in
the F, I, and S models. Note that $\Sigp$ is smaller near the
respective CR than the other shocked regions in all models.

The extent of spiral shocks depends sensitively on the pattern speed
of the spiral pattern as well as its strength. To produce
quasi-steady spiral shocks, the gas has to not only move faster than
the local sound speed relative to the perturbing potential, but also
have sufficient time to respond to one arm before encountering the
next arm. In Model F10G, the spiral shocks exist only at $R \simlt
\Rt= 17 \kpc$ where $\Mperp\simlt 12$ (Fig.~\ref{fig-mach}c). In
Models F05G and F20G, they extend up to $\Rt=15\kpc$ with
$\Mperp=10$ and to $\Rt=21\kpc$ with $\Mperp=15$, respectively. In
the I models, the termination radii of spiral shocks are $\Rt=22$,
$25$, and $32\kpc$ for $\F=5$, $10$, and $20\%$, respectively. The
time interval between two successive passages of the spiral
potential is $t_{\rm arm}=\pi/|\Omega-\Oms|$, while the arm-to-arm
crossing time of sound waves is $t_{\rm sound} = \pi R/\cs$. Thus
the condition for the formation of quasi-steady spiral shocks can be
written as
\begin{equation}\label{eq:t_ratio}
 \frac{t_{\rm
sound}}{t_{\rm arm}}=\frac{\Mperp}{\sin p_*} \simlt 20+100\F,
\end{equation}
for $5\%\leq \F\leq20\%$, insensitive to the arm pattern speed. If
this condition is not satisfied, the gas does not properly feel the
azimuthal variations of the imposed potential that is rotating too
rapidly.  This is consistent with the finding of \citet{bak74} that
gas with too large an entry velocity relative to the arms moves
almost ballistically, without producing a steady-state shock.  In
the S models, the entire simulation domain has $\Mperp/\sin
p_*\simlt 20$, and thus quasi-steady spiral shocks form over the
whole disk.

\begin{figure*}
\centering
\includegraphics[angle=0,width=0.98\textwidth]{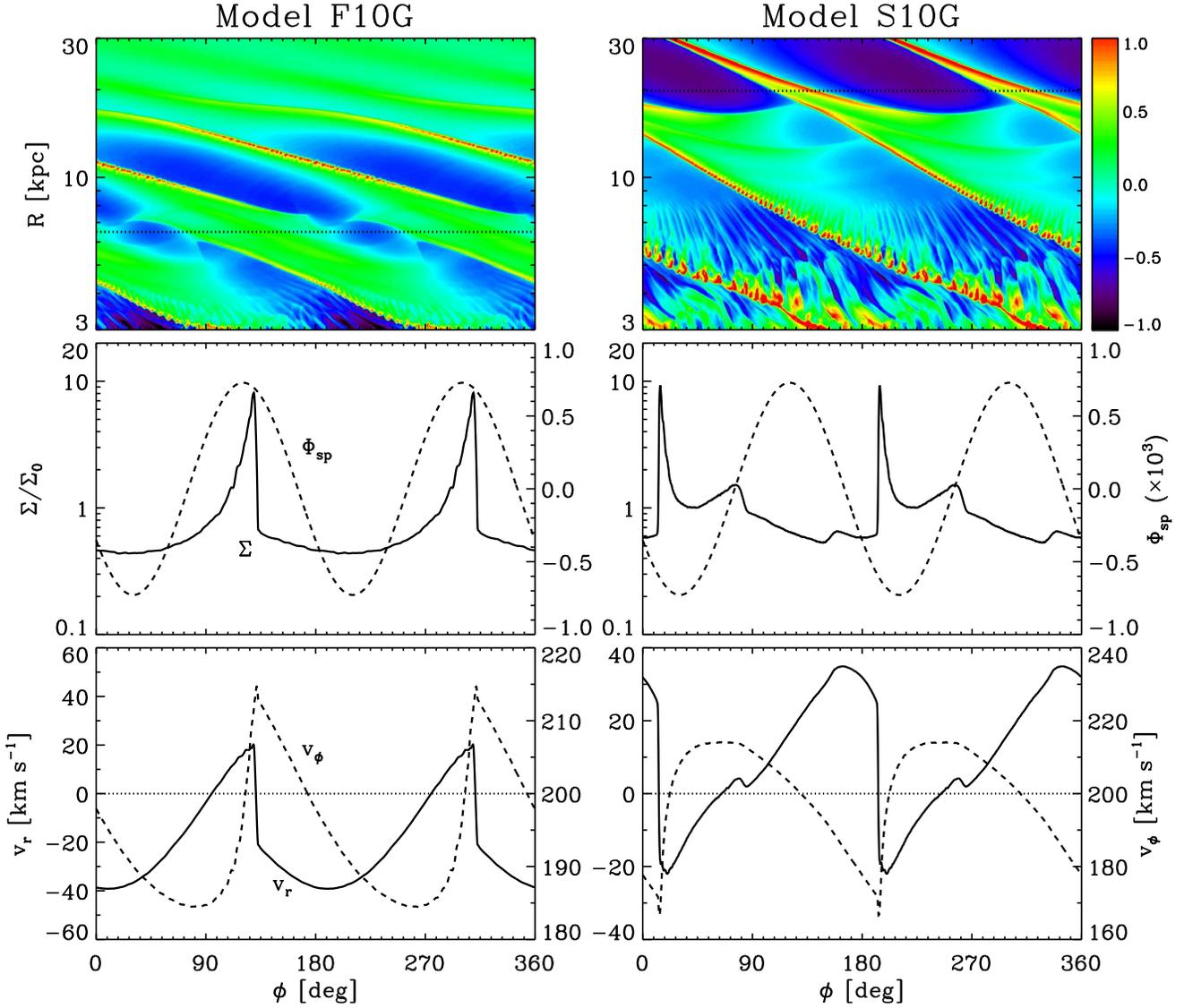}
\caption{Top: Logarithm of the gas surface density on the $\log
R$--$\phi$ plane for Models (left) F10G and (right) S10G at
$t=0.4\Gyr$. The CR of the arms is indicated by a horizontal dotted
line. Colorbar labels $\log(\Sigma/\Sigma_0)$. Middle: Azimuthal-cut
profiles of the surface density $\Sigma$ (solid; left $y$-axis) and
the spiral potential $\Psp$ (dashed; right $y$-axis) at $R=13\kpc$.
Bottom: The corresponding azimuthal profiles of the inertial-frame
radial (solid; left $y$-axis) and azimuthal (dashed; right $y$-axis)
velocities along the same cut. \label{fig-Phi}}
\end{figure*}

A conventional wisdom is that spiral shocks are absent in the CR
region where $\Mperp=0$. However, Figure \ref{fig-alpha}b reveals
that spiral shocks, albeit somewhat weak, are not completely absent
near the CR in the S models. There are two different kinds of spiral
shocks depending on the locations. The first kind, marked by plus
symbols, starting from the ILR is produced primarily by the gas
inside the CR that is rotating faster than the spiral potential. In
Model S10G, these spiral shocks become weaker with increasing $R$
owing to the decrease in $\Mperp$ and move across the CR located at
$\RCR=19.8\kpc$, extending all the way to the outer radial boundary.
In Models F10G and I10G, however, they barely extend across the
respective CR. The second kind, marked by squares, is generated by
the gas outside the CR rotating slower the pattern. In Model S10G,
they become weaker with decreasing $R$ from the outer boundary, move
inward across the CR, and cease to exist at $R\sim 17\kpc$. In Model
F10G and I10G, on the other hand, they are strongest at
$R\sim10\kpc$ and $15\kpc$, respectively, and do not extend inward
across the CRs. In what follows, we term the first and second spiral
shock the inner and outer shocks, respectively.

The extension of the spiral shocks across the CR in the S models is
caused by epicycle motions of perturbed gas elements near the CR by
the spiral potential. To illustrate this, we plot in Figure
\ref{fig-stm} instantaneous streamlines of the gas in Model S10G at
$t=0.4\Gyr$ in the frame corotating with the spiral potential. Only
a few streamlines around a Lagrangian point denoted by $L_1$ are
shown. The spiral potential induces radial velocity perturbations
with amplitude $\Delta v_R$ in the gas flows. In the local
approximation, the corresponding radial amplitude of the epicycle
orbits is $\Delta R = \Delta v_R/\kappa$ (e.g., \citealt{bin08}).
While a gas element originally located well inside the CR goes out
radially on the course of its epicycle motion, it meets a shock at
an oblique angle smaller than $90\degr$ due to fast rotation
relative to the pattern, and moves radially inward after the shock,
as exemplified by the streamline A.  But, gas elements located
closer to $L_1$ achieve larger epicycle phases, due to slower
relative rotation, when they hit the shocks.  Thus, the angles
between the incident streamlines and the shock fronts can become
larger than $90\degr$, causing the streamlines B--E to bend radially
outward after the shocks. These outwardly-moving gas elements
increasingly find themselves in the regions with smaller $\kappa$,
which in turn makes them move much farther than the original $\Delta
R$ implies. Consequently, the inner spiral shocks are smoothly
extended to the outer radial boundary in the S models. Note that the
dense arm gas outside the CR is bounded by two spiral shocks in
these models.

Similarly, the outer spiral shocks extend inward across the CR in
the S models, but in this case the radial excursion of the perturbed
elements is quite limited because they feel larger $\kappa$ as they
move inward. In Model S10G, for instance, the radial velocity
perturbations at $R=20\kpc$ is $\Delta v_R = 36\kms$. With
$\kappa=13\freq$ at this radius, the radial displacement is $\Delta
R=2.7\kpc$, which matches the numerical results well.  In the F and
I models, however, $\Delta R \sim 0.1\kpc$ due to lager $\kappa$,
which is too small a perturbation to make spiral shocks extended
across the CR.

Figure \ref{fig-Phi} plots the gas surface density in the $\log
R$--$\phi$ plane, together with the azimuthal cut profiles of
various quantities at $R=13\kpc$ for Models F10G and S10G at
$t=0.4\Gyr$. Relative to the spiral potential, the gas at this
radius in Model F10G (Model S10G) is moving in the negative
(positive) $\phi$-direction, forming shock fronts that are displaced
by $\sim 80\degr$ downstream ($\sim 20\degr$ upstream) from the
potential minima. The gas is compressed at the shock fronts,
enhancing the surface density there, while reducing the velocity
perpendicular to the shocks. The gas expands after the shock in
order to follow quasi-periodic galaxy rotation. The constraint of
the potential vorticity conservation requires the velocity parallel
to the shock to increase after the shock front, resulting in shear
reversal inside the gaseous arms with $\Sigma/\Sigma_0>2$ (e.g.,
\citealt{bal88,kim02}). Streaming velocities due to the spiral
shocks amount typically to $\sim40$--$60\kms$. The overall flow
pattern is similar to the observed profiles associated with spiral
arms in M51 (see, e.g., Fig.\ 6 of \citealt{she07}).

\subsubsection{Pitch Angle}\label{sec:pitch}

\begin{figure}
\centering
\includegraphics[angle=0,width=0.48\textwidth]{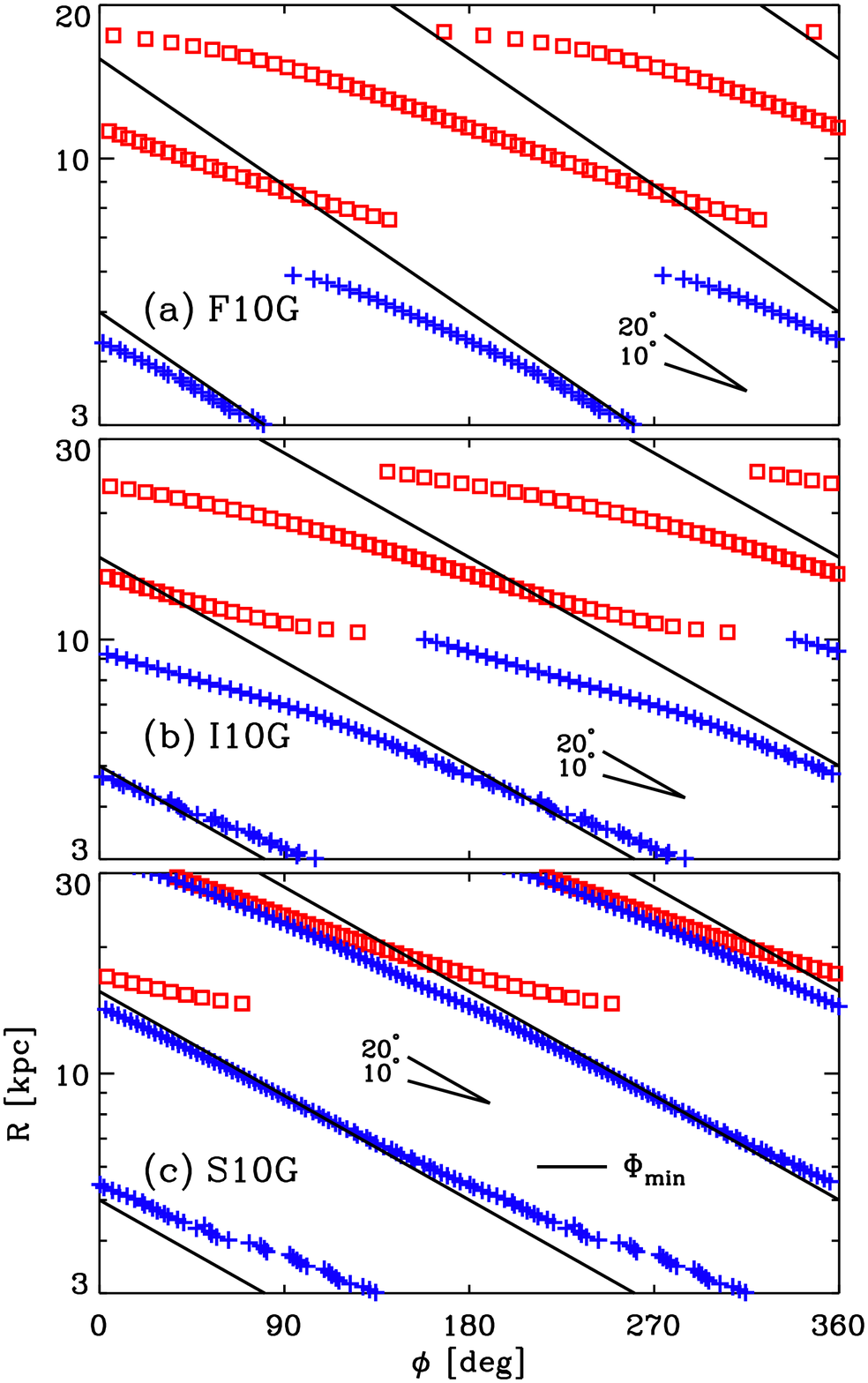}
\caption{Azimuthal positions of the inner (pluses) and outer
(squares) spiral shocks in comparison with the minimum loci (solid
lines) of the spiral-arm potential for Models (a) F10G, (b) I10G,
and (c) S10G at $t=0.4\Gyr$. Only the regions with $\alpha \geq 0.5$
are shown. In (a) and (b), the azimuthal locations of spiral shocks
change rapidly with $R$, resulting in a much smaller pitch angle
than the stellar arms, while the difference between the pitch angles
of the stellar and gaseous arms is small in (c).\label{fig-pitch}}
\end{figure}

Figure \ref{fig-pitch} compares the azimuthal positions of the inner
(pluses) and outer (squares) spiral shocks with the minima (solid
lines) of the external potential for our fiducial models at
$t=0.4\Gyr$. Only the regions with $\alpha \geq 0.5$ are shown. The
discontinuation of both inner and outer spiral shocks at the CR is
evident in Model F10G, while the inner spiral shocks extend all the
way to the outer boundary in Model S10G. The spiral shocks are
approximately logarithmic in shape over a wide range of radii, with
a pitch angle depending on $\Oms$. In Model S10G, the spiral shocks
are located very close to the potential minima and thus have a pitch
angle of $p_{\rm gas}\sim17\degr$, not much different from $p_*$. In
Models F10G and I10G, on the other hand, the shock positions deviate
considerably from the potential minima, resulting in much smaller
pitch angles of $p_{\rm gas}\sim 8\degr$ and $10\degr$,
respectively.

The dependence of the shock positions relative to the potential
minima is due to the tendency that stronger shocks form farther
downstream \citep{kim02}. As Figure \ref{fig-mach}c shows, spiral
shocks in Model S10G have $\Mperp\sim5$ at $R\sim7\kpc$ and are
located near the potential minima. As $R$ increases, $\Mperp$
decreases and the shocks become weaker, moving slightly toward the
upstream direction. In Model F10G, on the other hand, $\Mperp$
($\propto R$ for large $R$) varies a lot with $R$, leading to fairly
large variations in the shock positions. For instance, spiral shocks
at $R=8\kpc$ have $\Mperp=2$ and are placed near the potential
minima. At $R=12\kpc$, $\Mperp\sim7$ and the shocks are displaced by
$90\degr$ toward the downstream direction. At $R=17\kpc$ and beyond,
$\Mperp\simgt 12$ and no stationary configuration of spiral shocks
can be found in this model.

\begin{figure}
\centering
\includegraphics[angle=0,width=0.48\textwidth]{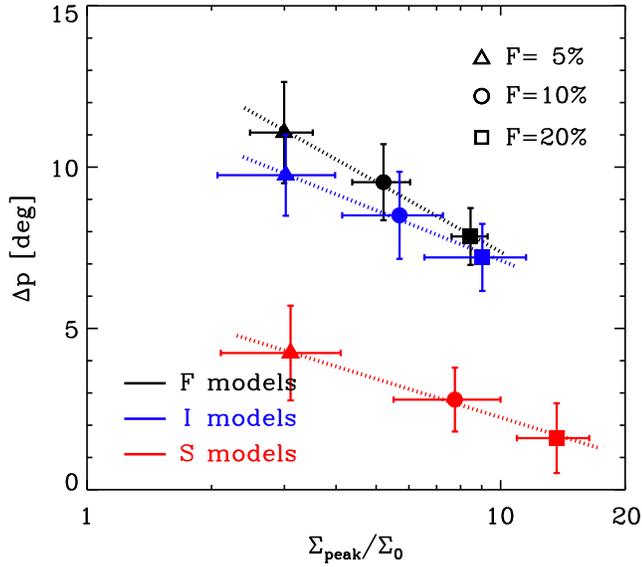}
\caption{Offsets, $\Delta p=p_*-p_{\rm gas}$, of the pitch angles
between the stellar and gaseous arms as a function of the peak shock
density $\Sigp$ averaged over $t=0.2$--$0.6\Gyr$ and $R=6$--$15\kpc$
for all self-gravitating models. Errorbars indicate the standard
deviations in $\Delta p$ and $\Sigp$. The dotted lines are our best
fits (Eq.\ [\ref{eq:delp}]).\label{fig-theta}}
\end{figure}

The radial dependence of $\Mperp$ makes $p_{\rm gas}$ smaller than
$p_*$. To quantify the offsets of the pitch angles, we define
$\Delta p \equiv p_* - p_{\rm gas}$, and plot them in Figure
\ref{fig-theta} as a function of the peak shock density $\Sigp$ for
various models with differing $\F$ and $\Oms$. Each symbol gives the
mean values of $\Delta p$ and $\Sigp$ averaged over
$t=0.2$--$0.6\Gyr$ and $R=6$--$15\kpc$, with errorbars indicating
the standard deviations. The dotted lines give our best fits
\begin{equation}\label{eq:delp}
\Delta p = \left\{
\begin{array}{ll}
15-7\log(\Sigp/\Sigma_0),\;\;\textrm{for F models}, \\
12-5\log(\Sigp/\Sigma_0),\;\;\textrm{for I models}, \\
\,\,\,6-4\log(\Sigp/\Sigma_0),\;\;\textrm{for S models}.
\end{array}\right.
\end{equation}
Since $\Delta p>0$ for a reasonable range of $\Sigp$, the pitch
angle of the gaseous arms puts the lower limit to that of the
stellar arms. In general, larger $\Sigp$ corresponds to smaller
$\Delta p$. Compared to the S models, models with larger $\Oms$ have
larger $\Delta p$ and smaller $\Sigp$. Models with larger $\F$ have
smaller $\Delta p$ and larger $\Sigp$ since a deeper spiral
potential tends to form shocks closer to the potential minima. That
the difference of $\Delta p$ between the F and I models is smaller
than that between the I and S models indicates that $\Delta p$ is
deeply related to $\RCR$.  This result suggests that one should be
cautious when inferring $p_*$ from $p_{\rm gas}$, especially when
$\Oms$ is large and $\F$ is small.

\section{Mass Drift}\label{sec:mdot}

The non-axisymmetric spiral pattern and the associated shocks are an
efficient means of angular momentum transport, causing gas elements
in orbital motions to move radially inward or outward depending on
the sign of $\Omega-\Oms$ (e.g., \citealt{shu92}). It is commonly
accepted that inside the CR where $\Omega > \Oms$, gas can lose angular
momentum from the spiral shocks, tending to move radially inward.
Outside the CR with $\Omega < \Oms$, on the other hand, spiral shocks
provide positive torque and thus cause the gas to move radially outward.
In this section, we quantify the rate of mass drift driven by spiral arms.

\begin{figure}
\centering
\includegraphics[angle=0,width=0.48\textwidth]{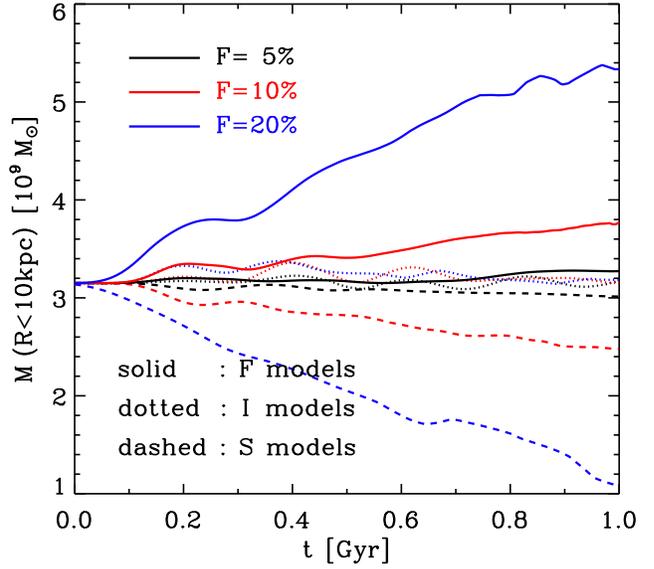}
\caption{Temporal variations of the integrated mass $M(R<10\kpc)$
within $10\kpc$ for all self-gravitating models. Solid lines are for
the S models exhibiting gas inflows, while dashed lines are for the
F models with gas outflows. For the I models, $M(R<10\kpc)$ plotted
as dotted lines does not change much over time. \label{fig-min}}
\end{figure}

\begin{figure}
\centering
\includegraphics[angle=0,width=0.48\textwidth]{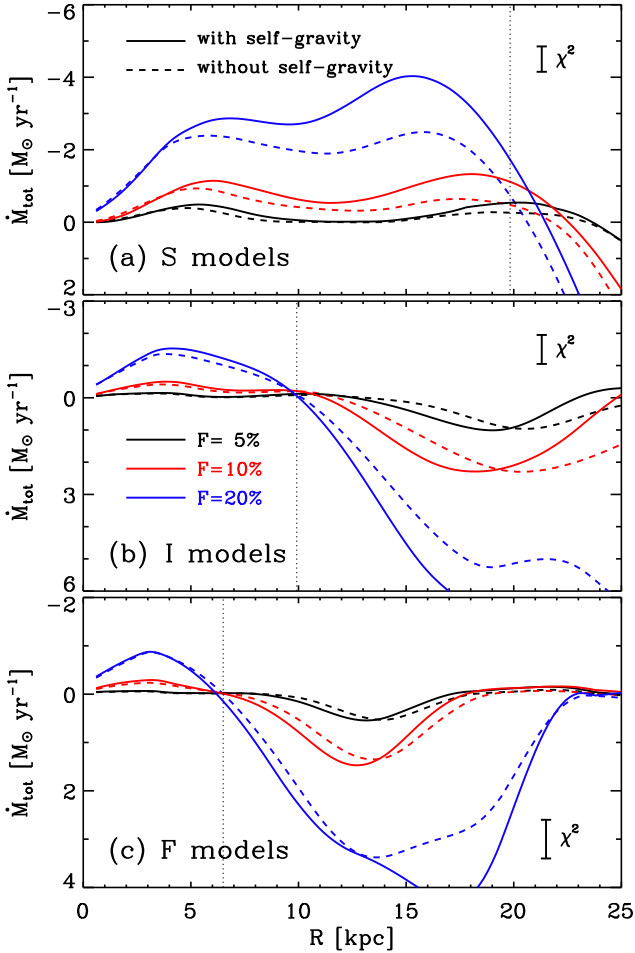}
\caption{Radial dependence of the mass drift rate $\Mtot\equiv
-dM(<R)/dt$ averaged over $t=0.2$--$0.8\Gyr$ for the (a) S (b) I,
and (c) F models. The solid and dashed lines represent
self-gravitating and non-self-gravitating models, respectively. The
short vertical bars marked by $\chi^2$ give the typical variations
of $\Mdot$ over time. The vertical dotted line in each panel
indicates the CR.\label{fig-mdot-arm}}
\end{figure}

To investigate the radial mass changes in our models, Figure
\ref{fig-min} plots temporal variations of the total gas mass, $M
(<10\kpc)$, within $R=10\kpc$ in various self-gravitating models.
The presence of the spiral potential makes $M(<10\kpc)$ increase
faster for models with stronger arms in the S models, while the F
models with large $\F$ exhibit decreases in $M(<10\kpc)$,
corresponding to mass outflows. The I models do not show any
noticeable secular changes in $M(<10\kpc)$ since the mass is
measured near the CR. Figure \ref{fig-mdot-arm} plots the radial
distributions of the mass drift rate, $\Mtot(R) \equiv -dM(<R)/dt$,
averaged over $t=0.2$--$0.8\Gyr$, for both self-gravitating (solid
lines) and non-self-gravitating (dashed lines) models. The vertical
bar marked by $\chi^2$ in each panel indicates the typical standard
deviations. Note that $\Mtot$ is negative for mass inflows and
positive for outflows. All the models show mass inflows inside the
CR, with the effect of self-gravity insignificant except for Model
S20G. In the S models, $\Mtot$ is relatively constant at $\sim -
(0.2$--$3)\Aunit$ over a wide range of $R$, which is significantly
larger than $\sim -(0.05$--$0.8)\Aunit$ and $\sim
-(0.1$--$1.2)\Aunit$ in the F and I counterparts, respectively,
owing to larger $\RCR$. In contrast, the regions outside the CR
clearly show mass outflows, with $\Mtot$ varying with $R$
considerably. Columns (2) and (3) of Table \ref{tbl-mdot} list
$\aMtot_{\rm in}$ and $\aMtot_{\rm out}$ averaged spatially over
$\RILR \simlt R \simlt \RCR$ and $\RCR\simlt R \simlt \Rt$,
respectively. Here, $\Rt$ represents the termination radius of the
spiral shocks for the F and I models, as mentioned in Section
\ref{sec:str}, and is taken to $35\kpc$ for the S models.

\begin{figure}
\centering
\includegraphics[angle=0,width=0.48\textwidth]{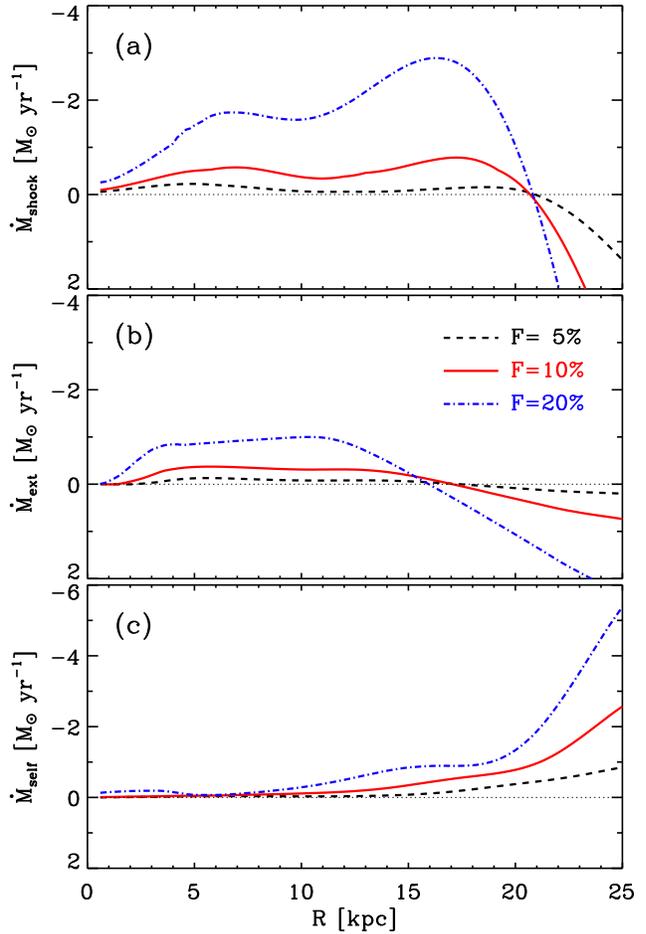}
\caption{Radial distributions of (a) $\Mshock$, (b) $\Mext$, and (c)
$\Mself$ for all self-gravitating, slow arm models, averaged over
$t=0.2$--$0.8\Gyr$. Models with stronger arms definitely have a
higher mass drift rate. Note that $\Mshock$ is negative inside the
CR and positive outside the CR. The self-gravitational contribution
which is always negative dominates at large $R$.
\label{fig-mdot-anal}}
\end{figure}

The radial gas drift in our models is caused by the combination of
three processes: (1) dissipation of angular momentum at spiral shocks,
(2) torque by the external spiral potential, and (3) torque by the
self-gravitational potential. The first two processes have previously
been well recognized by other authors (e.g.,
\citealt{kal72,rob72,lub86,hop11}), while the effect of self-gravity
did not receive much attention. We thus write
\begin{equation}\label{eq:Mflux}
\Mtot = \Mshock + \Mext + \Mself,
\end{equation}
where $\Mshock$, $\Mext$, and $\Mself$ denote the contributions of
spiral shocks, the external spiral potential, and the gaseous
self-gravity, respectively.  It is well known that $\Mext$ can be
expressed by
\begin{equation}\label{eq:Mext}
\Mext = \left(\frac{1}{R}\frac{\partial R^2\Omega} {\partial
R}\right)^{-1}\int_{-\pi}^{\pi} \Sigma \frac{\partial\Psp}{\partial
\phi}d\phi,
\end{equation}
(e.g., \citealt{lub86}). We similarly write the self-gravitational
contribution as
\begin{equation}\label{eq:Mself}
\Mself= \left(\frac{1}{R}\frac{\partial R^2\Omega} {\partial
R}\right)^{-1}\int_{-\pi}^{\pi} \Sigma \frac{\partial\Phi_{\rm
gas}}{\partial \phi}d\phi.
\end{equation}
There is no simple analytic expression for $\Mshock$.

\begin{figure}
\centering
\includegraphics[angle=0,width=0.48\textwidth]{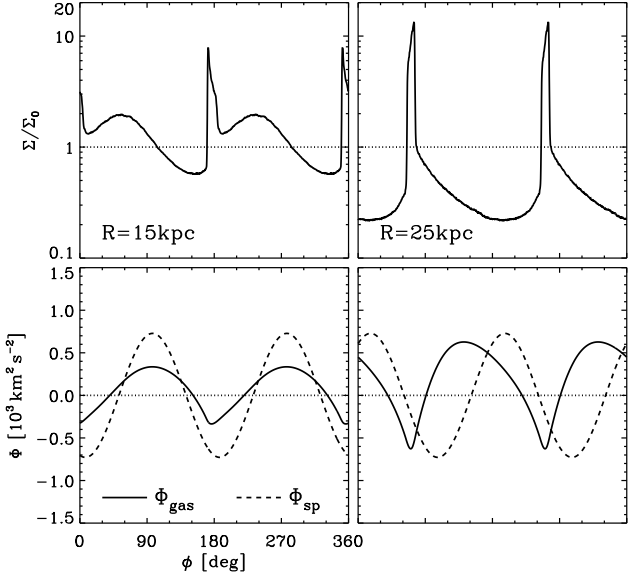}
\caption{Azimuthal distributions of (upper panels) gas surface
density and (lower panels) the spiral and self-gravitational
potentials at $R=15$ and $25\kpc$ for Model S10G when $t=0.4\Gyr$.
Note that gaseous arms outside the CR have larger density contrasts
and thus produce stronger self-gravitational forces than inside the
CR. \label{fig-cut}}
\end{figure}

\begin{table*}
\caption{Various Mass Drift Rates Induced by Spiral
Arms\label{tbl-mdot}}
\begin{center}
\begin{tabular}{ccccccccc}
\hline\hline Model & $\aMtot_{\rm in}$ & $\aMtot_{\rm out}$ &
$\aMshock_{\rm in}$ & $\aMshock_{\rm out}$ & $\aMext_{\rm in}$ &
$\aMext_{\rm out}$ &
$\aMself_{\rm in}$ & $\aMself_{\rm out}$ \\
(1) & (2) & (3) & (4) & (5) & (6) & (7) & (8) & (9) \\
\hline
F05G & $-0.054$ & $0.273$ & $-0.033$ & $0.181$ & $-0.018$ & $0.101$ & $-0.003$ & $-0.010$\\
F10G & $-0.249$ & $0.919$ & $-0.126$ & $0.518$ & $-0.112$ & $0.435$ & $-0.011$ & $-0.035$\\
F20G & $-0.781$ & $2.479$ & $-0.391$ & $1.134$ & $-0.274$ & $1.552$ & $-0.116$ & $-0.207$\\
I05G & $-0.095$ & $0.556$ & $-0.057$ & $0.414$ & $-0.032$ & $0.163$ & $-0.006$ & $-0.021$\\
I10G & $-0.361$ & $1.730$ & $-0.188$ & $1.153$ & $-0.153$ & $0.663$ & $-0.021$ & $-0.087$\\
I20G & $-1.184$ & $4.313$ & $-0.600$ & $2.415$ & $-0.489$ & $2.256$ & $-0.095$ & $-0.358$\\
S05G & $-0.205$ & $1.032$ & $-0.075$ & $1.024$ & $-0.079$ & $1.167$ & $-0.052$ & $-1.159$\\
S10G & $-0.876$ & $2.669$ & $-0.391$ & $2.185$ & $-0.302$ & $3.326$ & $-0.181$ & $-2.842$\\
S20G & $-3.074$ & $6.244$ & $-1.838$ & $3.429$ & $-0.799$ & $7.537$ & $-0.438$ & $-4.722$\\
\hline
F05N & $-0.046$ & $0.226$ & $-0.033$ & $0.110$ & $-0.012$ & $0.116$ & $0.0  $ & $0.0  $\\
F10N & $-0.203$ & $0.772$ & $-0.120$ & $0.321$ & $-0.083$ & $0.452$ & $0.0  $ & $0.0  $\\
F20N & $-0.778$ & $2.280$ & $-0.518$ & $0.802$ & $-0.260$ & $1.478$ & $0.0  $ & $0.0  $\\
I05N & $-0.080$ & $0.411$ & $-0.054$ & $0.221$ & $-0.026$ & $0.191$ & $0.0  $ & $0.0  $\\
I10N & $-0.292$ & $1.322$ & $-0.178$ & $0.602$ & $-0.114$ & $0.720$ & $0.0  $ & $0.0  $\\
I20N & $-1.116$ & $3.315$ & $-0.694$ & $1.199$ & $-0.423$ & $2.115$ & $0.0  $ & $0.0  $\\
S05N & $-0.159$ & $0.665$ & $-0.111$ & $0.024$ & $-0.048$ & $0.641$ & $0.0  $ & $0.0  $\\
S10N & $-0.608$ & $2.889$ & $-0.404$ & $0.764$ & $-0.204$ & $2.125$ & $0.0  $ & $0.0  $\\
S20N & $-2.240$ & $6.952$ & $-1.552$ & $1.489$ & $-0.688$ & $5.463$ & $0.0  $ & $0.0  $\\
\hline
\end{tabular}
\end{center}
\medskip
{Note. -- All values are averaged over $0.2\Gyr\leq t \leq 1.0\Gyr$
and in units of $\Msun\yr^{-1}$. Columns (2), (4), (6), and (8) are
the mass drift rates averaged over $\RILR\simlt R\simlt \RCR$, while
Columns (3), (5), (7), and (9) are the values averaged over from the
CR to the termination radius of the spiral shocks.}
\end{table*}

Using equations (\ref{eq:Mext}) and (\ref{eq:Mself}), we calculate
$\Mext$ and $\Mself$ from our numerical results, and then $\Mshock$
from equation (\ref{eq:Mflux}). Figure \ref{fig-mdot-anal} plots the
radial distributions of $\Mshock$, $\Mext$, and $\Mself$ from the
self-gravitating, slow arm models. These values averaged over $\RILR
\simlt R\simlt \RCR$ or $\RCR\simlt R \simlt \Rt$ are listed in
Columns (4)--(9) of Table \ref{tbl-mdot} for all the models. As
expected, $\Mshock$ is negative in most of the region inside the CR
and positive outside the CR, although $\Mshock=0$ does not
correspond exactly to the CR. This is because the inner and outer
spiral shocks extend across the CR, as explained in Section
\ref{sec:str}. Since the effect of the inner shocks is stronger on
the mass drift than that of the outer shocks, $\Mshock=0$ occurs at
$R\sim 21\kpc$, roughly independent of $\F$, slightly outside the
CR. Well inside the CR, $\Mext$ is negative mostly due to the torque
on the gas in the postshock expanding zones rather than on the gas
at the shock fronts. Although the latter has the highest density,
its contribution to $\Mext$ turns out to be insignificant, except
near the CR, since it is located very close to the potential minima
in the S models. Near the CR, $\Mext$ becomes positive due to the
torque on the gas at the shock fronts that are displaced
substantially from the potential minima toward the upstream
direction.

The self-gravitational torque overwhelms the other torques outside
the CR. This is because self-gravity becomes relatively more
important at larger $R$, as equation (\ref{eq:toomre}) suggests. In
addition, gaseous arms in outer regions are bounded by two spiral
shocks and thus relatively thick. Figure \ref{fig-cut} compares the
azimuthal distributions of gas surface density together with the
spiral and self-gravitational potentials at (left) $R=15\kpc$ and
(right) $25\kpc$ in Model S10G at $t=0.4\Gyr$. The shock-bounded
arms outside the CR form at the expense of lower interarm density
and thus have stronger self-gravitational forces, compared to those
inside the CR. Note that the density distribution at $R=25\kpc$ is
slightly asymmetric with respect to the minima of $\Pgas$, with
larger density at the side with $\partial\Pgas/\partial\phi<0$. The
corresponding self-gravitational torque is thus negative over most
of the simulation domain.

A stronger spiral potential leads to larger $|\Mtot|$. Our numerical
results for the mass drift for all self-gravitating models can be
fitted as
\begin{equation}\label{eq:Mtot2}
\aMtot_{\rm in} \approx\left\{
\begin{array}{ll}
-4\F (0.2 + 4\F)(\Sigma/\Sigma_0)\,, \;\;\textrm{for F models}, \\
-5\F (0.2 + 5\F)(\Sigma/\Sigma_0)\,, \;\;\textrm{for I models}, \\
-7\F (0.2 + 10\F)(\Sigma/\Sigma_0)\,,\;\;\textrm{for S models}, \\
\end{array}\right.
\end{equation}
and
\begin{equation}\label{eq:Mtot3}
\aMtot_{\rm out} \approx\left\{
\begin{array}{ll}
5\F (1 + 7\F)(\Sigma/\Sigma_0)\,, \;\;\textrm{for F models}, \\
5\F (2 + 11\F)(\Sigma/\Sigma_0)\,, \;\;\textrm{for I models}, \\
5\F (4 + 11\F)(\Sigma/\Sigma_0)\,,\;\;\textrm{for S models}, \\
\end{array}\right.
\end{equation}
in units of $\Aunit$, both of which are accurate within
$\sim0.1\Aunit$ for $0.05\leq \F \leq 0.2$. In these models, the
angular momentum loss at spiral shocks, the external gravitational
torque, and the self-gravitational torque account for about $50\%$,
$40\%$, and $10\%$ of the total, on average, respectively, roughly
independent of $\F$. In the S models, the corresponding radial
inflow velocity is $v_d = \Mtot/(2\pi R \Sigma_0) \sim 1 \kms$ at
$R=10\kpc$, with the associated timescale comparable to the Hubble
time.

\section{Line-of-Sight Velocity}\label{sec:los}

\begin{figure*}
\centering
\includegraphics[angle=0,width=0.98\textwidth]{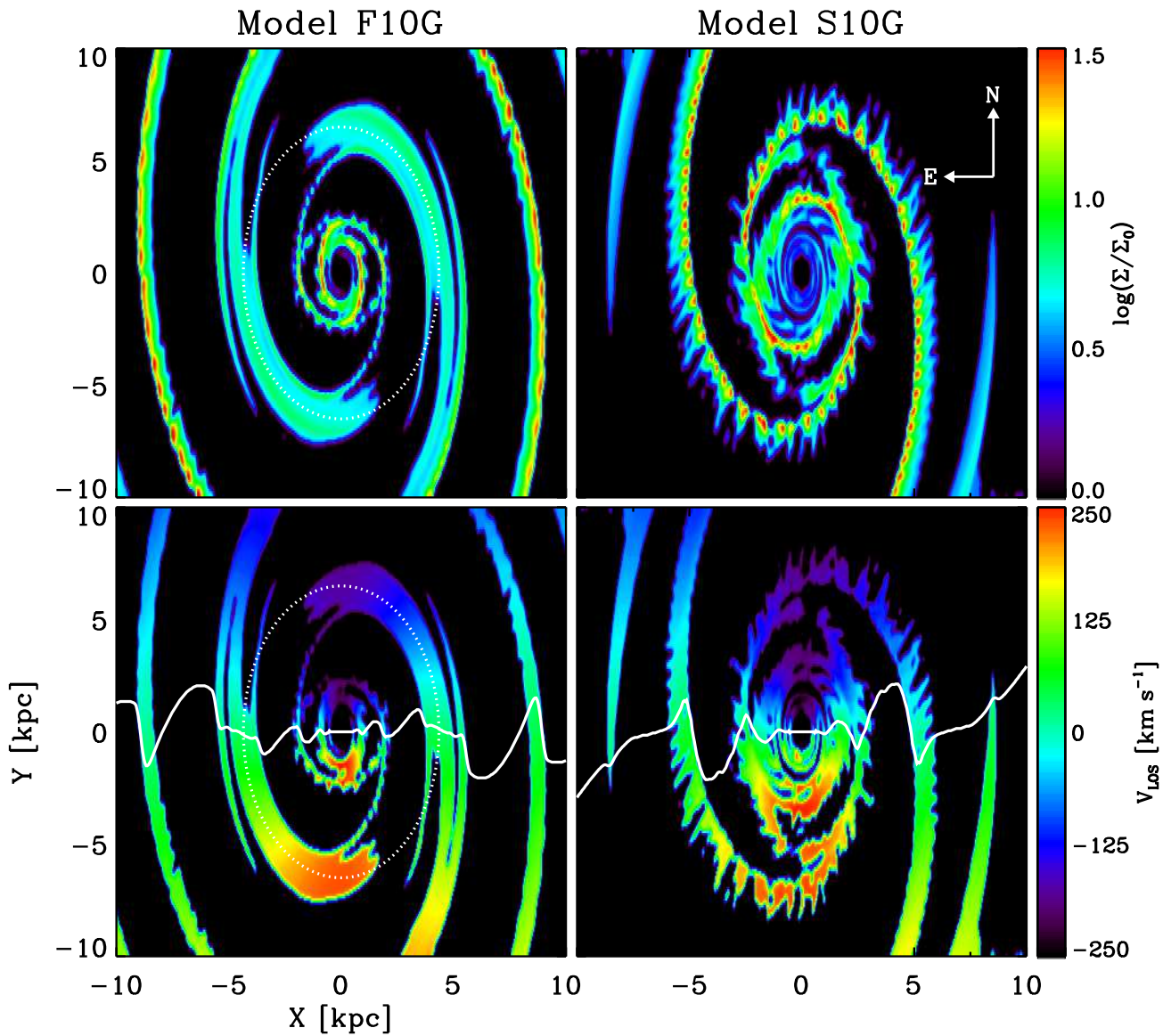}
\caption{Synthetic, projected maps of (upper panels) gas surface
density $\Sigma$ in logarithmic scale and (lower panels) the
line-of-sight velocity $\Vlos$ for Models (left) F10G and (right)
S10G at $t=0.4\Gyr$.  Only the regions with $\Sigma/\Sigma_0\ge1$
are shown. The inclination angle of the disk is set to $65\degr$
relative to the plane of the sky, and the position angle of the arms
at $R=5\kpc$ is taken to be $-25\degr$ from the north. The dotted
ellipses in the left panels indicate the location of the CR in Model
F10G. The solid curves in the lower panels draw the loci of
$\Vlos=0$. Upper and lower colorbars label $\log(\Sigma/\Sigma_0)$
and $\Vlos$ in units of km s$^{-1}$,
respectively.\label{fig-vlosmod}}
\end{figure*}

The distribution of line-of-sight velocities in the plane of sky can
be a useful diagnostic tool in analyzing the gas kinematics in disk
galaxies. Figure \ref{fig-vlosmod} plots the synthetic, projected
maps of (upper panels) the gas surface density in logarithmic
scales, and (lower panels) the line-of-sight velocity $\Vlos$ for
(left) Models F10G and (right) S10G at $t=0.4\Gyr$. Only the regions
with $\Sigma/\Sigma_0\ge 1$ are shown to mimic strong radio
emissions from overdense regions. The galaxy is rotating in the
counterclockwise direction. In each panel, the disk is assumed to be
inclined arbitrarily by $65\degr$ with respect to the plane of sky
($XY$-plane) such that west of the disk is the near side. We choose
$-25\degr$ as the position angle of the arms at $R=5\kpc$ measured
from the north (positive $Y$-axis). The dotted ovals in the left
panels indicate the location of the CR in Model F10G, while the
solid curves in the lower panels draw the loci of $\Vlos=0$. Density
and velocity data are smoothed by a Gaussian beam with a width of
$0.12\kpc$. The colorbars label $\log(\Sigma/\Sigma_0)$ and $\Vlos$
in units of $\kms$.

In the central regions at $R\simlt2\kpc$, the zero velocity curve
runs almost parallel to the nodal line ($X$-axis) for both models.
As $R$ increases, however, the pattern speed makes several
differences in the arm morphologies and the distribution of $\Vlos$
that can possibly be discerned observationally. First, gaseous arms
are more tightly wound in Model F10G due to a rapid change in
$\Mperp$ than in Model S10G. Second, spiral arms located outside the
CR are bounded by shock fronts at the outer edges, while those
inside the CR are shocked at the inner edges. Consequently, the gas
density in the arms is distributed asymmetrically along a radial cut
such that it is larger at smaller $R$ outside the CR and at larger
$R$ inside the CR, as the upper panels of Figure \ref{fig-vlosmod}
illustrate. Third, gas streaming motions associated with spiral
shocks amount to $\sim40$--$60\kms$, which can make the gas rotate
slower than the pattern even inside the CR (see Fig.\
\ref{fig-Phi}). Thus the zero velocity curve near the western
(eastern) arm in Model S10G strongly bends downward (upward), toward
the opposite direction to the galaxy rotation. On the other hand,
the gas rotates faster than the pattern outside the CR, which causes
the zero velocity curve to bend upward near the arms in the western
parts of the disk, as in Model F10G. These differences can be used
to determine whether observed segments of the arms are located
inside or outside the CR. We will discuss this further in
application to NGC 3627 in Section \ref{sec:dis}.

\section{Summary and Discussion}\label{sec:sum_dis}

\subsection{Summary}\label{sec:sum}

We have presented the results of grid-based hydrodynamic simulations
on spiral structures and radial mass drift in disk galaxies driven
by spiral arms. The gaseous disk is assumed to be infinitesimally
thin, unmagnetized, isothermal with the sound speed of $\cs=10\kms$,
and initially uniform with surface density $\Sigma_0=10\Surf$. For
the spiral arms, we impose a rigidly-rotating logarithmic
gravitational potential with pitch angle $p_*=20\degr$, strength
$\F$, and pattern speed $\Oms$. To study the dependence of the shock
structure and the mass drift rates on $\F$ and $\Oms$, we consider
three types of models in which the arm is rotating fast at
$\Oms=30\freq$, intermediately at $\Oms=20\freq$, or slow at
$\Oms=10\freq$, which are referred to as the F, I, and S model,
respectively. We also vary $\F$ between 5 and 20\%. The main results
of this paper are summarize as follows.

1. \emph{Extent of Spiral shocks.} -- The radial extent of spiral
shocks depends rather sensitively on the arm pattern speed. In the F
and I models, spiral shocks exist only up to $\Rt\sim 17\kpc$ and
$\sim25\kpc$, respectively, while the outer region is almost
featureless other than weak trailing waves. This is because when
equation (\ref{eq:t_ratio}) is not fulfilled, gas perturbed by one
arm does not have sufficient time to adjust itself to the imposed
spiral potential before encountering the next arm. That is, the
rapid rotation of the potential makes itself effectively smoothed
considerably along the azimuthal direction, and gas moves almost
ballistically \citep{bak74}. In the S models, on the other hand, the
whole disk satisfies $\Mperp/\sin p_*\simlt 20$ and forms spiral
shocks across the entire simulation domain. In these models with a
slow pattern, spiral shocks are not terminated at the CR due to
epicycle motions of perturbed gas elements. Since a gas element on
its epicycle orbit achieves larger (smaller) amplitudes as it moves
radially outward (inward), the spiral shocks produced inside the CR
are able to extend all the way to the outer radial boundary, while
those originally formed outside the CR extend only slightly inward
of the CR.  As a consequence, the dense arm gas outside the CR is
bounded by two spiral shocks in the S models.

2. \emph{Relation between the pitch angle and shock strength.} -- In
a quasi-steady state, stronger spiral shocks tend to form at farther
downstream relative to the minima of the imposed spiral potential.
Since $\Mperp$ varies systematically with $R$, this makes the pitch
angle $p_{\rm gas}$ of the gaseous arms smaller than that of the
stellar arms. In our models, the offset between $p_{\rm gas}$ and
$p_*$ amounts to $\sim2\degr$--$12\degr$, and is larger for smaller
$\F$ since a deeper potential tends to have shocks closer to its
minima. It is also larger for models with larger $\Oms$ due to
larger radial variations of $\Mperp$.  Equation (\ref{eq:delp})
gives our fits to $\Delta p=p_* - p_{\rm gas}$ against the peak
shock density $\Sigp$ of the gaseous spiral arms.

3. \emph{Mass Drift} -- The non-axisymmetric spiral potential is an
efficient means of angular momentum transport, initiating radial
drift of gas that would otherwise be in circular motions. In our
models, the radial mass drift is caused by a combination of three
processes: angular momentum loss at spiral shocks, external
gravitational torque, and self-gravitational torque. While
self-gravitational torque is always negative, it is usually smaller
than the other torques inside the CR.  On the other hand, the
direction of the mass drift by the shock loss and external torque
depends on the sign of $\Omega-\Oms$, such that it is radially
inward inside the CR and outward outside the CR. The resulting mass
inflow rate, averaged over $\RILR\simlt R\simlt \RCR$, is in the
range $\aMtot_{\rm in}\sim -(0.05$--$3.0)\Aunit$, with a larger
value corresponding to stronger and/or slower arms, as described by
equation (\ref{eq:Mtot2}). The shock loss and external spiral
potential account for about 50\% and 40\% of the total,
respectively.

4. \emph{Line-of-Sight Velocity} -- Since the spiral arms cause
streaming motions in the gas flows whose amplitudes depend on the
arm pattern speed, the related line-of-sight velocity $\Vlos$ and
the density distribution across the arms can potentially provide
information on the arm pattern speed. Gaseous arms located outside
the CR have a larger density at larger $R$ along a radial cut. In
this case, the gas in the arms rotates faster than the pattern due
to the streaming motions, tending to make the locus of $\Vlos=0$
bend in the same way as the direction of galaxy rotation. In
contrast, gaseous arms located inside the CR have a larger density
at smaller $R$, and the $\Vlos=0$ curve near the arms bends in the
opposite sense to the galaxy rotation (see Fig.\ \ref{fig-vlosmod}).

\begin{figure*}
\centering
\includegraphics[angle=0,width=0.98\textwidth]{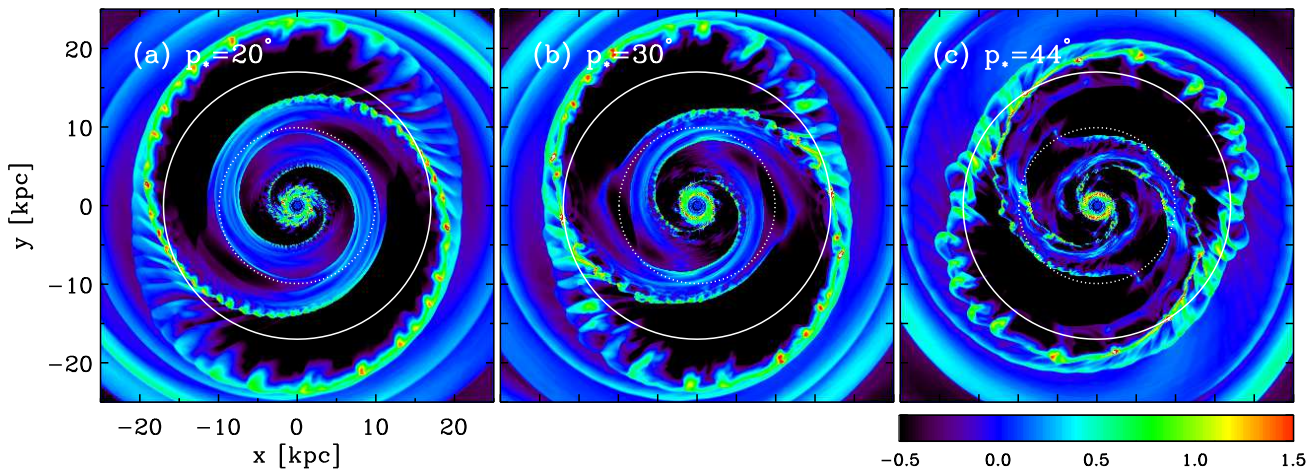}
\caption{Logarithm of gas surface density at $t=1.0\Gyr$ for
self-gravitating models with pitch angle (a) $p_*=20\degr$, (b)
$p_*=30\degr$, and (c) $p_*=44\degr$.  The other parameters are the
same as in Model I10G. In each panel, the dotted and solid circles
draw the CR and OLR.  A ring-like structure forms near the position
where the outer end of a gaseous arm curls radially in to be
connected to the other arm. Colorbar labels ${\rm
log}(\Sigma/\Sigma_0)$. \label{fig-res}}
\end{figure*}

\subsection{Discussion}\label{sec:dis}

Our result that the radial extent of gaseous arms depends on the arm
pattern speed is overall consistent with the finding of
\citet{pat94} who showed using SPH simulations that gaseous spirals
exist up to the CR when $\Oms$ is large ($\sim37.4\freq$), while
they extend to the end of the stellar spiral arms when $\Oms$ is
small ($\sim12.5\freq$). \citet{pat94} further found that an oval
ring forms, possibly by the resonance, near the OLR area.  A similar
ring-like structure forms in the outer regions of Model I10G  at
late time when the outer ends of trailing gaseous arms curve back
radially inward to touch the other arms (see Fig.\ \ref{fig-sigt}).
The main difference in the model parameters between our and their
models is the pitch angle of the stellar potential. To explore the
effect of $p_*$ on the position of ring-like structures that form,
we have run two additional models with $p_*=30\degr$ and $44\degr$,
while the other parameters remain identical to Model
I10G.\footnote{Models in \citet{pat94} took $p_*=44\degr$.} Figure
\ref{fig-res} plots the distributions of surface density at
$t=1\Gyr$ from these additional runs as well as Model I10G. In each
panel, the dotted and solid circles mark the CR and OLR,
respectively. An oval-like structure forms in all models, although
its size tends to decrease with increasing $p_*$. Model I10G with
$p_*=20\degr$ has an oval-like structure well outside the OLR, while
it is coincidentally at the OLR. Since the arm termination (and thus
the ring formation) occurs at the radius where equation
(\ref{eq:t_ratio}) is satisfied, one can expect a smaller ring when
$p_*$ is larger, fully consistent with our numerical results. This
demonstrates that the formation of a ring-like structure near or
beyond the OLR is not due to the resonance.

It is interesting to apply our results to the spatial extent of
observed arms in the barred-spiral galaxy M83. In this galaxy, the
spiral pattern rotates relatively rapidly at $\Oms\approx 45\freq$
and has a pitch angle of $p_*\approx 16\degr$ (e.g.,
\citealt{lor91,zim04}). The rotational velocity at the flat part is
$\sim180\kms$ \citep{lun04}, so that the CR is located at
$\sim4\kpc$, corresponding to $\sim3^\prime$ at the distance of
$4.5\Mpc$ \citep{thm03}. The radio data of \citet{lun04} show that
the gaseous arms are weaker at the CR than at the neighboring arms,
similarly to our results shown in Figure \ref{fig-mach}. The arms
extend up to $\sim 6^\prime$, while there is a plenty of gas with
$\Sigma\sim2\Surf$ in the outer regions where the gaseous arms are
absent \citep{cro02,lun04}. Although the termination radius of the
gaseous arms in M83 is close to the OLR, it is uncertain whether the
OLR plays a central role in limiting the arm extent.  We note that
the radius of $6^\prime$ corresponds to $\Mperp/\sin p_*\sim 23$--30
for the observed CO velocity dispersions of $\cs\sim7.8\pm0.9\kms$
\citep{lun04}, suggesting that the idea of arm termination by too
large $\Mperp$ is not inconsistent with the observed gaseous arms in
M83 with $\F\sim5$--10\%.

To measure the mass drift rate unaffected by a radial density
gradient for given $\F$, we have employed simple disk models with
radially constant $\Sigma_0$ and $\F$.  In models with a slow patten
speed, this inevitably results in readily discernable spiral shocks
all the way to the outer radial boundary. In reality, however, gas
surface density in spiral galaxies appears to drop off exponentially
or more rapidly (e.g., \citealt{big12}). In addition, the stellar
spiral potential is likely to become shallower with increasing $R$
beyond the CR (e.g., \citealt{con86,con88,pat91}). Although
arm-to-interam density contrasts are likely unchanged by the
background density (especially when self-gravity is unimportant),
small values of $\Sigma_0$ and $\F$ would make it difficult to
detect gaseous spiral arms at large radii in real spiral galaxies.

The tendency of spiral shocks moving toward the upstream direction
with increasing $R$ was reported by \citet{git04}, and our results
further show that the displacement of spiral shocks is larger when
$\Oms$ is larger. This is consistent with the results of
\citet{pat94} who found that gaseous arms are much tighter than the
stellar pattern in their high-$\Oms$ models. \citet{git04} also
noted that in addition to stellar and gaseous arms, there are
star-forming arms traced by \HII regions, all of which may
have different pitch angles such that $p_* > p_{\rm gas} > p_{\rm
SF}$ if the time offset between the gaseous and star-forming arms is
independent of $R$. Indeed, \citet{gro98} showed that the stellar
arms traced by $K^\prime$-band observations are more loosely wound
than the optical arms for a sample of five galaxies. Although
\citet{dav12} more recently found that the arm pitch angles in the
optical band are almost identical to those in the near-IR band
within observational uncertainties (see also \citealt{sei06}), their
fitting results plotted in their Figure 13 still show that the
latter is larger by $\sim2\degr$ than the former for $p_* \sim
10\degr$--$30\degr$, consistent with our numerical results.
\citet{mart12} also reported that arm pitch angles are larger in the
$H$-band than in the $B$-band. More accurate observational estimates
are required to explore the dependence on $\Oms$ of the offsets
between arm pitch angles in different wavelength bands.

In application to the gas accretion in the Milky Way, \citet{lub86}
used a local model of spiral shocks by considering the reaction of
stellar waves back to the gaseous self-gravity and calculated the
mass inflows rate due to the spiral shocks and external
gravitational torque.  They found that the total inflow rate is in
the range $\Mdot\sim-(0.2$--$0.4)\Aunit$ for $\F=3\%$, consistent
with the extrapolation of our numerical results. They also found
that the external potential is about three times more effective than
the viscous torque, which is different from out finding that the
effect of the spiral shocks in removing angular momentum is slightly
larger than that of the external potential. This difference is
presumably caused by the fact that \citet{lub86} included physical
viscosity accounting for cloud collisions. This smears out the
shocks and moves the peak density toward downstream (see also
\citealt{kim08,kim10}), which tends to enhance the external torque
and reduce the angular-momentum loss at the shock fronts.

\begin{figure}
\centering
\includegraphics[angle=0,width=0.48\textwidth]{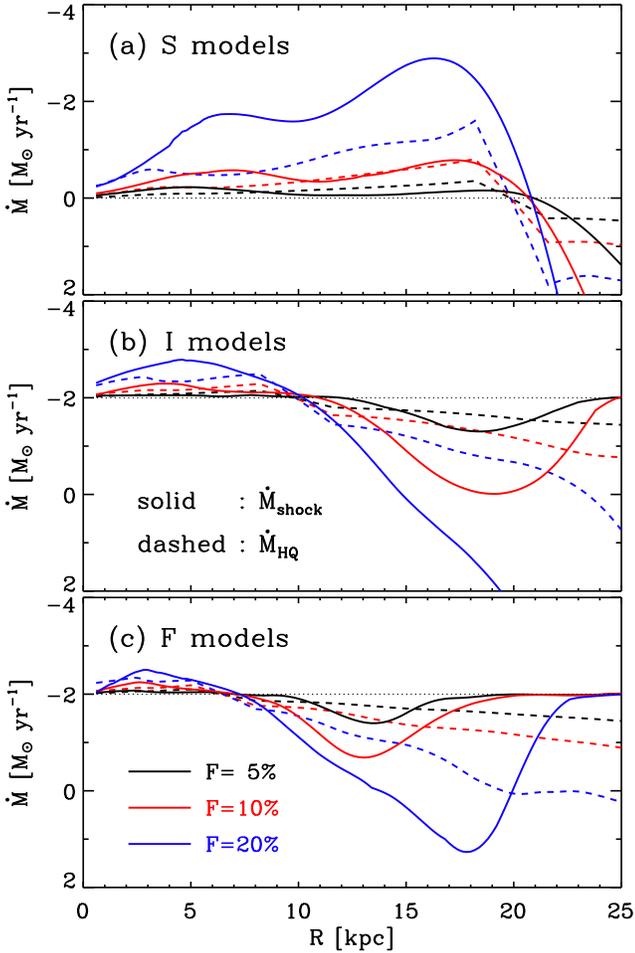}
\caption{Comparison of the mass inflow rate $\Mshock$ (solid lines)
by angular-momentum loss at spiral shocks in our models with the
analytic estimate $\Mdot_{\rm HQ}$ of \citet{hop11} (dashed lines)
given by equation (\ref{eq:HQ}) for the (a) S, (b), I, and (c) F
models. Note that $\Mdot_{\rm HQ}$ agrees approximately with
$\Mshock$ only in the S models with small $\F$, while it differs
considerably from $\Mshock$ for the small-$\Oms$ with large $\F$ and
all the large-$\Oms$ models.\label{fig-hq}}
\end{figure}

\citet{hop11} put forward a simple analytic model for angular
momentum transport and related gas inflows driven by a
non-axisymmetric stellar potential. By neglecting the effects of
thermal pressure as well as the flow velocity relative to the
potential, they considered orbit crossing of collisionless particles
as a criterion for shock formation and derived the mass inflow rate
\begin{equation}\label{eq:HQ}
\Mdot_{\rm HQ} = -\bar\Sigma R^2 \Omega \F\tan p_* {\rm
sign}(\Omega-\Oms) |f(\zeta)|,
\end{equation}
where $\bar\Sigma$ is the azimuthally-averaged gas density and
$f(\zeta)$ denotes the correction factor of order unity for the
degree of the orbit crossing, $\zeta$. They confirmed that equation
(\ref{eq:HQ}) is in good agreement with their numerical results when
bar-like stellar modes dominate the perturbations. To check whether
equation (\ref{eq:HQ}) is also good for the cases dominated by a
spiral potential, Figure \ref{fig-hq} compares $\Mdot_{\rm HQ}$
(dashed lines) with $\Mshock$ (solid lines) from our numerical
simulations. Note that equation (\ref{eq:HQ}) can be a good
approximation for the mass inflows due to the shock loss
\emph{alone}, only for weak spiral shocks in the S models in which
shock fronts are located very close to the potential minima. On the
other hand, equation (\ref{eq:HQ}) fails to describe the mass drift
accurately for the slow-arm models with large $\F$ and for the F
models where spiral shocks are displaced significantly from the
potential minima. In this case, the shock formation requires
consideration of thermal pressure as well as incident velocities,
which were neglected in \citet{hop11}.

By analyzing Sloan Digital Sky Survey DR7 data, \citet{com09} found
that about 20\% of 266 galaxies with measured bar strength host dust
lanes. On the other hand, numerical simulations with only a bar
potential without spiral arms show that dust lanes remain strong
only for $0.1\Gyr$ around the time when the bar potential achieves
its full strength (e.g., \citealt{kim12b,kim12}). This implies that
barred galaxies with strong dust lanes should be either dynamically
young or supplied with fresh gas. Our numerical results in this
paper suggest that spiral arms can be efficient to transport gas
from outside to the central region provided that the spiral arms
have quite a low pattern speed so as to have a large CR radius.

In addition to enhancing the mass in the central region, mass
inflows by spiral arms also may help to increase the rate of star
formation occurring in the nuclear ring induced by a bar potential.
\citet{seo13} numerically found that without spiral arms or gas
infall from halo, the star formation rate in nuclear rings decays to
small values below $\sim 1 \Aunit$ after showing a strong burst that
lasts only about $0.1\Gyr$. This is in contrast to the claim that
star formation in nuclear rings is a long-lived phenomenon
($\sim1$--$2\Gyr$), with multiple episodes (e.g.,
\citealt{all06,sar07,van13}). In Model S10G, gas flows radially
inward at a rate $\Mtot \sim -1.0\Aunit$ at $R=5\kpc$, which would
fuel star formation on the galactic center when enough mass is
accumulated to undergo gravitational collapse. It would be
interesting to study how much star formation is enhanced in nuclear
rings by an addition of outer spiral arms, which will direct our
future research.

By employing a simple isothermal equation of state, our models
produce spiral shocks with quite a large density contrast between
arm and interarm regions.  For instance, \citet{she07} reported that
spiral arms seen in CO observations of M51 typically have an
arm-to-interarm density contrast of $\sim5$, while it is $\sim 20$
in our models (see, e.g., Fig.\ \ref{fig-Phi}). Even considering the
beam smearing of the CO data, the density contrast in our models
seems to be larger than observed values. In real galaxies, there are
many physical processes including magnetic fields and star formation
that can affect spiral-shock gas dynamics considerably.  Magnetic
fields that are pervasive in the interstellar medium (e.g.,
\citealt{bec96,fle11}) can make the isothermal gas behave as if it
were adiabatic with index of 2 for one-dimensional compression
(e.g., \citealt{shu92}), reducing the arm-peak density substantially
(e.g., \citealt{kim02,kos02,lee12}). Star formation and ensuing
feedback occurring inside spiral arms is able to provide the arm gas
with turbulent kinetic energy which not only tends to disperse dense
gas but also triggers new star formation (e.g.,
\citealt{mk04,mck07,she12,kim13}).  A reduced arm density may
decrease the mass drift rate, while magnetic fields themselves can
be a source of additional angular momentum transport via tension
forces. Therefore, more realistic quantitative assessment of the
mass drift rate caused by spiral arms requires consideration of
these processes as well as radiative heating and cooling.

Finally, we discuss the distribution of gas density and $\Vlos$ in
the arms found in our models in comparison with observations.
\citet{reg02} presented a velocity map of CO emission from the
barred-spiral galaxy NGC 3672, derived from Gaussian fits to the
line profiles. Their Figure 6 shows that the observed distributions
of $\Vlos$ in the outer regions, especially the loci of $\Vlos=0$,
match those in Model S10G better than in Model F10G presented in
Figure \ref{fig-vlosmod}, suggesting that the spiral arms in NGC
3672 are located inside the their CR. A close inspection of Figure
4a of \citet{reg02} (see also Figure 1c of \citealt{che03}) reveals
that the density profiles of the gaseous arms along a radial cut
have a steeper gradient at the inner edge than the outer edge, which
also hints that the spiral arms rotate slow. Since the CR of a
strong bar is placed just outside the bar end, these all indicate
that the spiral arms in NGC 3672 have a lower pattern speed than the
bar. Indeed, \citet{ran04} used the method of \citet{tre84} to show
that the bar in NGC 3627 has a pattern speed of
$\Omb=50^{+3}_{-8}\freq$, while the southern extension of the
western spiral arm has $\Oms=23\pm4\freq$. The corresponding CR
radii of the bar and arms are at $R=1.3^\prime$ and $3.3^\prime$,
respectively, indicating that the observed spiral arms in
\citet{reg02} are located in between the CRs of the bar and spiral
arms, consistent with kinematic features in our models.

\section*{Acknowledgments}

We gratefully acknowledge helpful discussions with Bruce G.\
Elmegreen and Eve C. Ostriker. We are also grateful to the referee
for an insightful and constructive report. This work was supported
by the National Research Foundation of Korea (NRF) grant funded by
the Korean government (MEST), No.\ 2010-0000712. The computation of
this work was supported by the Supercomputing Center/Korea Institute
of Science and Technology Information with supercomputing resources
including technical support (KSC-2012-C3-19). Hospitality at APCTP
during the 7th Korean Astrophysics Workshop is kindly acknowledged.

\bsp
\label{lastpage}


\begin{thebibliography}{}
\bibitem[\protect\citeauthoryear{Allard et al.}{2006}]{all06}
  Allard E.~L., Knapen J.~H., Peletier R.\ F., \& Sarzi M.\  2006,
  MNRAS, 371, 1087
\bibitem[\protect\citeauthoryear{Amaral \& L\'epine}{1997}]{ama97}
  Amaral L.~H., \& L\'epine J.~R.~D.\ 1997, MNRAS, 286, 885
\bibitem[\protect\citeauthoryear{Ann \& Thakur}{2005}]{ann05}
  Ann H.\ B., \& Thakur P.\ 2005, ApJ, 620, 197
\bibitem[\protect\citeauthoryear{Artymowicz \& Lubow}{1992}]{art92}
  Artymowicz P., \& Lubow S.\ H.\ 1992, ApJ, 389, 129
\bibitem[\protect\citeauthoryear{Athanassoula}{1992}]{ath92}
  Athanassoula E.\ 1992, MNRAS, 259, 345
\bibitem[\protect\citeauthoryear{Athanassoula}{2002}]{ath02}
  Athanassoula E.\ 2002, ApJ, 569, L83
\bibitem[\protect\citeauthoryear{Baker \& Barker}{1974}]{bak74}
  Baker P.~L., \& Barker P.~K.\ 1974, A\&A, 36, 179
\bibitem[\protect\citeauthoryear{Balbus}{1988}]{bal88}
  Balbus S.~A.\ 1988, ApJ, 324, 60
\bibitem[\protect\citeauthoryear{Beck}{1996}]{bec96}
  Beck R.\ 1996, ARAA, 34, 155
\bibitem[\protect\citeauthoryear{Bertin \& Lin}{1996}]{ber96}
  Bertin G., \& Lin C.~C.\ 1996, Spiral Structure in Galaxies: A Density Wave
  Theory (Cambridge: MIT Press)
\bibitem[\protect\citeauthoryear{Bertin et al.}{1989a}]{ber89a}
  Bertin G., Lin C.~C., Lowe S.~A., \& Thustans R.~P.\
  1989a, ApJ, 338, 78
\bibitem[\protect\citeauthoryear{Bertin et al.}{1989b}]{ber89b}
  Bertin G., Lin C.~C., Lowe S.~A., \& Thustans R.~P.\
  1989b, ApJ, 338, 104
\bibitem[\protect\citeauthoryear{Bigiel \& Blitz}{2012}]{big12}
  Bigiel F., \& Blitz L.\ 2012, ApJ, 756, 183
\bibitem[\protect\citeauthoryear{Binney \& Tremaine}{2008}]{bin08}
  Binney J., \& Tremaine S.\ 2008, Galactic Dynamics
  (2nd ed.; Princeton, NJ: Princeton Univ.\ Press)
\bibitem[\protect\citeauthoryear{Bournaud \& Combes}{2002}]{bou02}
  Bournaud F., \& Combes F.\ 2002, A\&A, 392, 83
\bibitem[\protect\citeauthoryear{Brunetti et al.}{2011}]{bru11}
  Brunett M., Chiappini C., \& Pfenniger D.\ 2011, A\&A, 534, 44
\bibitem[\protect\citeauthoryear{Buta}{2013}]{but13}
  Buta R.\ 2013, Secular Evolution of Galaxies: XXIII Canary
  Islands Winter School of Astrophysics, eds.\ J.\ Falcon-Barroso,
  \& J.\ Knapen (Cambridge: Cambridge University Press),
  in press; arXiv:1304.3529
\bibitem[\protect\citeauthoryear{Buta \& Combes}{1996}]{but96}
  Buta R., \& Combes F.\ 1996, Fund.\ Cosmic Phys., 17, 95
\bibitem[\protect\citeauthoryear{Chakrabarti et al.}{2003}]{cha03}
  Chakrabarti S., Laughlin G., \& Shu F.~H.\ 2003, ApJ, 596, 220
\bibitem[\protect\citeauthoryear{Chemin et al.}{2003}]{che03}
  Chemin L., Cayatte V., Balkowski C., Marcelin M., Amram P.,
  van Driel W., \& Flores H.\ 2003, A\&A, 405, 89
\bibitem[\protect\citeauthoryear{Colella \& Woodward}{1984}]{col84}
  Colella P., \& Woodward P. R. 1984, Comput. Phys., 54, 174
\bibitem[\protect\citeauthoryear{Comer\'on et al.}{2009}]{com09}
  Comer\'on S., Mart\'inez-Valpuesta I., Knapen J.\ H., \&
  Beckman J.\ E.\ 2009, ApJ, 706, L256
\bibitem[\protect\citeauthoryear{Comer\'on et al.}{2010}]{com10}
  Comer\'on S., Knapen J.\ H., Beckman J.\ E., Laurikainen E., Salo H.,
  Mart\'inez-Valpuesta \& Buta R.\ J.\ 2010, MNRAS, 402, 2462
\bibitem[\protect\citeauthoryear{Contopoulos \& Grosb\o l}{1986}]{con86}
  Contopoulos G., \& Grosb\o l P.\ 1986, A\&A, 155, 11
\bibitem[\protect\citeauthoryear{Contopoulos \& Grosb\o l}{1988}]{con88}
  Contopoulos G., \& Grosb\o l P.\ 1986, A\&A, 197, 83
\bibitem[\protect\citeauthoryear{Crosthwaite et al.}{2002}]{cro02}
  Crosthwaite L.~P., Turner J.~L., Buchholz L., Ho P.~T.~P., \&
  Martin R.~N.\ 2002, AJ, 123, 1892
\bibitem[\protect\citeauthoryear{Davis et al.}{2012}]{dav12}
  Davis B.~L., Berrier J.~C., Shields D.~W., et al.\ 2012, ApJS,
  199, 33
\bibitem[\protect\citeauthoryear{Dobbs \& Bonnell}{2006}]{dob06}
  Dobbs C.~L., \& Bonnell I.~A. 2006, MNRAS, 367, 873
\bibitem[\protect\citeauthoryear{Dobbs et al.}{2011}]{dob11}
  Dobbs C.~L., Burkert A., \& Pringle J.~E.\ 2011, MNRAS, 417, 1318
\bibitem[\protect\citeauthoryear{Donner \& Thomasson}{1994}]{don94}
  Donner K.~J., \& Thomasson M.\ 1994, A\&A, 290, 785
\bibitem[\protect\citeauthoryear{Dwarkadas \& Balbus}{1996}]{dwa96}
  Dwarkadas V.~V., \& Balbus S.~A.\ 1996, ApJ, 467, 87
\bibitem[\protect\citeauthoryear{Elmegreen}{1995}]{elm95}
  Elmegreen B.~G.\ 1995, Molecular Clouds and Star Formation, ed.
  C.~Yuan \& H.~You (Singapore: World Scientific), p.~149
\bibitem[\protect\citeauthoryear{Englmaier \& Gerhard}{1997}]{eng97}
  Englmaier P., \& Gerhard O.\ 1997, MNRAS, 287, 57
\bibitem[\protect\citeauthoryear{Fathi et al.}{2009}]{fat09}
  Fathi K., Beckman J.~E., Pi\~{n}ol-Ferre N., Hernandez O.,
  Mart\'inez-Valpuesta I., \& Carignan C.\ 2009, ApJ, 704, 1657
\bibitem[\protect\citeauthoryear{Fletcher et al.}{2011}]{fle11}
  Fletcher A., Beck R., Shukurov A., Berkhuijsen E.~M., \&
  Horellou C.\ 2011, MNRAS, 412, 2396
\bibitem[\protect\citeauthoryear{Foyle et al.}{2010}]{foy10}
  Foyle K., Rix H.-W., \& Zibetti S.\ 2010, MNRAS, 407, 2010
\bibitem[\protect\citeauthoryear{Garc\'{i}a-Burillo et al.}{2009}]{gar09}
  Garc\'{i}a-Burillo S., Fern\'{a}ndez-Garc\'{i}a S., Combes F.,
  et al.\ 2009, A\&A, 496, 85
\bibitem[\protect\citeauthoryear{Gittins \& Clarke}{2004}]{git04}
  Gittins D.~M., \& Clarke C. J. 2005, MNRAS, 349, 909
\bibitem[\protect\citeauthoryear{G\'omez \& Cox}{2002}]{gom02}
  G\'omez G.~C., \& Cox D.~P.\ 2002, ApJ, 580, 235
\bibitem[\protect\citeauthoryear{G\'omez et al.}{2013}]{gom13}
  G\'omez G.~C., Pichardo B., \& Martos M.~A.\ 2013, MNRAS, 430, 3010
\bibitem[\protect\citeauthoryear{Grosb$\o$l \& Patsis}{1998}]{gro98}
  Grosb$\o$l P.~J., \& Patsis P.~A.\ 1998, A\&A, 336, 840
\bibitem[\protect\citeauthoryear{Haan et al.}{2009}]{haa09}
  Haan S., Schinnerer E., Emsellem E., Garc\'{i}a-Burillo S.,
  Combes F., Mundell C.~G., \& Rix H.-W.\ 2009, ApJ, 692, 1623
\bibitem[\protect\citeauthoryear{Hanawa \& Kikuchi}{2012}]{han12}
  Hanawa T., \& Kikuchi D.\ 2012, ASP Conference Series V.\ 459:
  Numerical Modeling of Space Plasma Flows: ASTRONUM-2011, eds.\
  N.~V.~Pogorelov, J.~A.~Font, E.~Audit, \& G.~P.~Zank (ASP: San Francisco),
  p.~310
\bibitem[\protect\citeauthoryear{Heller \& Shlosman}{1994}]{hel94}
  Heller C. H., \& Shlosman I. 1994, ApJ, 424, 84
\bibitem[\protect\citeauthoryear{Hopkins \& Quataert}{2011}]{hop11}
  Hopkins P., \& Quataert E.\ 2011, MNRAS, 415, 1027
\bibitem[\protect\citeauthoryear{Hunt et al.}{2008}]{hun08}
  Hunt L.~K., et al.\ 2009, ApJ, 482, 133
\bibitem[\protect\citeauthoryear{Jogee et al.}{2005}]{jog05}
  Jogee S., Scoville N., \& Kenney J.~D.~P.\ 2005, ApJ, 630, 837
\bibitem[\protect\citeauthoryear{Kalnajs}{1971}]{kal71}
  Kalnajs A.~J.\ 1971, ApJ, 166, 275
\bibitem[\protect\citeauthoryear{Kalnajs}{1972}]{kal72}
  Kalnajs A.~J.\ 1972, ApJ, 11, L41
\bibitem[\protect\citeauthoryear{Knapen et al.}{2000}]{kna00}
  Knapen J.~H., Shlosman I., \& Peletier R.~F.\ 2000, ApJ, 529, 93
\bibitem[\protect\citeauthoryear{Kim et al.}{2008}]{kim08}
  Kim C.-G., Kim W.-T., \& Ostriker E.~C.\ 2008, ApJ, 681, 1148
\bibitem[\protect\citeauthoryear{Kim et al.}{2010}]{kim10}
  Kim C.-G., Kim W.-T., \& Ostriker E.~C.\ 2010, ApJ, 720, 1454
\bibitem[\protect\citeauthoryear{Kim et al.}{2013}]{kim13}
  Kim C.-G., Ostriker E.~C., \& Kim W.-T.\ \ 2013, ApJ, 776, 1
\bibitem[\protect\citeauthoryear{Kim \& Ostriker}{2002}]{kim02}
  Kim W.-T., \& Ostriker E.~C. 2002, ApJ, 570, 132
\bibitem[\protect\citeauthoryear{Kim \& Ostriker}{2006}]{kim06}
  Kim W.-T., \& Ostriker E.~C. 2006, ApJ, 646, 213
\bibitem[\protect\citeauthoryear{Kim \& Ostriker}{2007}]{kim07}
  Kim W.-T., \& Ostriker E.~C. 2007, ApJ, 660, 1232
\bibitem[\protect\citeauthoryear{Kim et al.}{2002}]{kos02}
  Kim W.-T., Ostriker E.~C., \& Stone J.~M.\ 2002, ApJ, 581, 1080
\bibitem[\protect\citeauthoryear{Kim \& Stone}{2012}]{kim12}
  Kim W.-T., \& Stone J.~M. 2012, ApJ, 751, 124
\bibitem[\protect\citeauthoryear{Kim et al.}{2012a}]{kim12a}
  Kim W.-T., Seo W.-Y., Stone J.~M., Yoon D., \& Teuben P.~J. 2012,
  ApJ, 747, 60
\bibitem[\protect\citeauthoryear{Kim et al.}{2012b}]{kim12b}
  Kim W.-T., Seo W.-Y., \& Kim Yonghwi 2012b, ApJ, 758, 14
\bibitem[\protect\citeauthoryear{Kormendy \& Kennicutt}{2004}]{kor04}
  Kormendy J., \& Kennicutt R.~C.\ 2004, ARA\&A, 42, 603
\bibitem[\protect\citeauthoryear{Lacey \& Fall}{1985}]{lac85}
  Lacey C.~G., \& Fall S.~M.\ 1985, ApJ, 290, 154
\bibitem[\protect\citeauthoryear{Laurikainen et al.}{2004}]{lau04}
  Laurikainen E., Salo H., Buta R., \& Vasylyev S.\ 2004, MNRAS, 355, 1251
\bibitem[\protect\citeauthoryear{Lee \& Shu}{2012}]{lee12}
  Lee W.-K., \& Shu F.~H.\ 2012, ApJ, 756, 45
\bibitem[\protect\citeauthoryear{Lin \& Lau}{1979}]{lin79}
  Lin C.~C., \& Lau Y.~Y.\ 1979, St.~A.~M., 60, 97
\bibitem[\protect\citeauthoryear{Lin \& Shu}{1964}]{lin64}
  Lin C.~C., \& Shu F.~H.\ 1964, ApJ, 140, 646
\bibitem[\protect\citeauthoryear{Lin \& Shu}{1966}]{lin66}
  Lin C.~C., \& Shu F.~H.\ 1966, Proc.~Natl.~Acad.~Sci., 55, 229
\bibitem[\protect\citeauthoryear{Lord \& Kenney}{1991}]{lor91}
  Lord S.~D., \& Kenney J.~D.~P.\ 1991, ApJ, 381, 130
\bibitem[\protect\citeauthoryear{Lubow et al.}{1986}]{lub86}
  Lubow S.~H., Balbus S.~A., \& Cowie L.~L.\ 1986, ApJ, 309, 496
\bibitem[\protect\citeauthoryear{Lundgren et al.}{2004}]{lun04}
  Lundgren A.~A., Olofsson H., Wiklind T., \& Rydbeck G.\ 2004,
  A\&A, 422, 865
\bibitem[\protect\citeauthoryear{Lynden-Bell \& Kalnajs}{1972}]{lyn72}
  Lynden-Bell D., \& Kalnajs A.~J.\ 1972, MNRAS, 157, 1
\bibitem[\protect\citeauthoryear{Maciejewski}{2004}]{mac04}
  Maciejewski W.\ 2004, MNRAS, 354, 892
\bibitem[\protect\citeauthoryear{Mac Low \& Klessen}{2004}]{mk04}
  Mac Low M., \& Klessen R.~S.\ 2004, RvMP, 76, 125
\bibitem[\protect\citeauthoryear{Mart\'inez-Garc\'ia}{2012}]{mart12}
  Mart\'inez-Garc\'ia E.~E.\ 2012, ApJ, 744, 92
\bibitem[\protect\citeauthoryear{Mart\'inez-Garc\'ia et al.}{2011}]{mart11}
  Mart\'inez-Garc\'ia E.~E.,
  \& Gonz\'alez-L\'opezlira R.~A.\ 2011, ApJ, 734, 122
\bibitem[\protect\citeauthoryear{Martos et al.}{2004}]{mar04}
  Martos M., Hernandez X., Y\'a\~nez M., Moreno E., \& Pichardo B.\ 2004,
  MNRAS, 350, L47
\bibitem[\protect\citeauthoryear{McKee \& Ostriker}{2007}]{mck07}
  McKee C.~F., \& Ostriker E.~C.\ 2007, ARA\&A, 45, 565
\bibitem[\protect\citeauthoryear{Oh et al.}{2008}]{oh08}
  Oh S.~H., Kim W.-T., Lee H.~M., \& Kim J.\ 2008, ApJ, 683, 94
\bibitem[\protect\citeauthoryear{Patsis et al.}{1991}]{pat91}
  Patsis P.~A., Contopoulos G., \& Grosb\o l P.\ 1991, A\&A, 243, 373
\bibitem[\protect\citeauthoryear{Patsis et al.}{1994}]{pat94}
  Patsis P.~A., Hiotelis N., Contopoulos G., \& Grosb\o l P.\
  1994, A\&A, 286, 46
\bibitem[\protect\citeauthoryear{Patsis et al.}{1997}]{pat97}
  Patsis P.~A., Grosb\o l P., \& Hiotelis N.\ 1997, A\&A, 323, 762
\bibitem[\protect\citeauthoryear{Patsis \& Athanassoula}{2000}]{pat00}
  Patsis P.~A., \& Athanassoula E.\ 2000, A\&A, 358, 45
\bibitem[\protect\citeauthoryear{Piner et al.}{1995}]{pin95}
  Piner B.~G., Stone J.~M., \& Teuben P.~J.\ 1995, ApJ, 449, 508
\bibitem[\protect\citeauthoryear{Rand \& Wallin}{2004}]{ran04}
  Rand R.~J., \& Wallin J.~F.\ 2004, ApJ, 614, 142
\bibitem[\protect\citeauthoryear{Rautiainen \& Salo}{1999}]{rau99}
  Rautiainen P., \& Salo H.\ 1999, A\&A, 348, 737
\bibitem[\protect\citeauthoryear{Regan \& Mulchaey}{1999}]{reg99}
  Regan M.~W., \& Mulchaey J.~S.\ 1999, AJ, 117, 2676
\bibitem[\protect\citeauthoryear{Regan et al.}{2002}]{reg02}
  Regan M.~W., Sheth K., Teuben P.~J., \& Vogel S.~N.\ 2002, ApJ, 574, 126
\bibitem[\protect\citeauthoryear{Roberts}{1969}]{rob69}
  Roberts W.~W.\ 1969, ApJ, 158, 123
\bibitem[\protect\citeauthoryear{Roberts \& Shu}{1972}]{rob72}
  Roberts W.~W., \& Shu F.~H.\ 1972, ApJ, 12, 49
\bibitem[\protect\citeauthoryear{Roca-F\`{a}brega et al.}{2013}]{roc13}
  Roca-F\`{a}brega S., Valenzuela O., Figueras F., Romero-G\'{o}mez M.,
  Vel\'{a}zquez H., Antoja T., \& Pichardo B.\ 2013, MNRAS, 432, 2878
\bibitem[\protect\citeauthoryear{Ro\v{s}kar et al.}{2008}]{ros08}
  Ro\v{s}kar R., Debattista V.~P., Quinn T.~R.,
  Stinson G.~S., \& Wadsley J.\ 2008, ApJ, 684, L79
\bibitem[\protect\citeauthoryear{Sanders \& Huntley}{1976}]{san76}
  Sanders R.~H., \& Huntley J.~M.\ 1976, ApJ, 209, 53
\bibitem[\protect\citeauthoryear{Sarzi et al.}{2007}]{sar07}
  Sarzi M., Allard E.\ L, Knapen J.\ H., \& Mazzuca L.\ M.\ 2007,
  MNRAS, 380, 949
\bibitem[\protect\citeauthoryear{Seigar et al.}{2006}]{sei06}
  Seigar M.~S., Bullock J.~S., Barth A.~A., \& Ho L.~C.\ 2006,
  ApJ, 645, 1012
\bibitem[\protect\citeauthoryear{Sellwood}{2011}]{sel11}
  Sellwood J.~A.\ 2011, MNRAS, 410, 1637
\bibitem[\protect\citeauthoryear{Sellwood}{2013}]{sel13}
  Sellwood J.~A.\ 2013, to appear in Reviews of Modern Physics; arXiv:1310.0403
\bibitem[\protect\citeauthoryear{Sellwood \& Binney}{2002}]{sel02}
  Sellwood J.~A., \& Binney J.~J.\ 2002, MNRAS, 336, 785
\bibitem[\protect\citeauthoryear{Sellwood \& Sparke}{1988}]{sel88}
  Sellwood J.~A., \& Sparke L.~S.\ 1988, MNRAS, 231, 25
\bibitem[\protect\citeauthoryear{Seo \& Kim}{2013}]{seo13}
  Seo W.-Y., \& Kim W.-T.\ 2013, ApJ, 769, 100
\bibitem[\protect\citeauthoryear{Shetty \& Ostriker}{2006}]{she06}
  Shetty R., \& Ostriker E.~C.\ 2006, ApJ, 647, 997
\bibitem[\protect\citeauthoryear{Shetty \& Ostriker}{2008}]{she08}
  Shetty R., \& Ostriker E.~C.\ 2008, ApJ, 684, 978
\bibitem[\protect\citeauthoryear{Shetty \& Ostriker}{2012}]{she12}
  Shetty R., \& Ostriker E.~C.\ 2012, ApJ, 754, 2
\bibitem[\protect\citeauthoryear{Shetty et al.}{2007}]{she07}
  Shetty R., Vogel S.~N., Ostriker E.~C., \& Teuben P.~J.\ 2007,
  ApJ, 665, 1138
\bibitem[\protect\citeauthoryear{Shlosman et al.}{1990}]{shl90}
  Shlosman I., Begelman M.~C., \& Frank J.\ 1990, Nature, 345, 679
\bibitem[\protect\citeauthoryear{Shu}{1992}]{shu92} Shu F. H. 1992,
  The Physics of Astrophysics. II. Gas Dynamics (Mill Valley: Univ. Science Books)
\bibitem[\protect\citeauthoryear{Slyz et al.}{2003}]{sly03}
  Slyz A.~D., Kranz T., \& Rix H.-W.\ 2003, MNRAS, 346, 1162
\bibitem[\protect\citeauthoryear{Thakur et al.}{2009}]{tha09}
  Thakur P., Ann H.~B., \& Jiang I.\ 2009, ApJ, 693, 586
\bibitem[\protect\citeauthoryear{Thim et al.}{2003}]{thm03}
  Thim F., Tammann G.~A., Saha A., et al.\ 2003, ApJ, 590, 256
\bibitem[\protect\citeauthoryear{Toomre}{1964}]{too64}
  Toomre A.\ 1964, ApJ, 139, 1217
\bibitem[\protect\citeauthoryear{Toomre}{1981}]{too81}
  Toomre A.\ 1981, in Structure and Evolution
  of Normal Galaxies, ed.~S.~M.~Fall \& D.~Lynden-Bell
  (Cambridge: Cambridge Univ.\ Press), p.~111
\bibitem[\protect\citeauthoryear{Tremaine \& Weinberg}{1984}]{tre84}
  Tremaine S., \& Weinberg M.~D.\ 1984, ApJ, 282, L5
\bibitem[\protect\citeauthoryear{van der Laan et al}{2013}]{van13}
  van der Laan T.~P.~R., Schinnerer E., Emsellem E., et al.\
  2013, A\&A, 551, 81
\bibitem[\protect\citeauthoryear{van de Ven \& Fathi}{2010}]{van10}
  van de Ven G., \& Fathi K.\ 2010, ApJ, 723, 767
\bibitem[\protect\citeauthoryear{Wada}{2008}]{wad08}
  Wada K.\ 2008, ApJ, 675, 188
\bibitem[\protect\citeauthoryear{Wada \& Koda}{2004}]{wad04}
  Wada K., \& Koda J.\ 2004, MNRAS, 349, 270
\bibitem[\protect\citeauthoryear{Wada et al.}{2011}]{wad11}
  Wada K., Baba J., \& Saitoh T.~R.\ 2011, ApJ, 735, 1
\bibitem[\protect\citeauthoryear{Y\'{a}\~{n}ez et al.}{2008}]{yan08}
  Y\'{a}\~{n}ez M.~A., Norman M.~L., Martos M.~A., \& Hayes J.~C.\
  2008, ApJ, 672, 207
\bibitem[\protect\citeauthoryear{Zhang}{1996}]{zha96}
  Zhang X.\ 1996, ApJ, 457, 125
\bibitem[\protect\citeauthoryear{Zimmer et al.}{2004}]{zim04}
  Zimmer P., Rand R.~J., \& McGraw J.~T.\ 2004, ApJ, 607, 285
\end{thebibliography}
\end{document}